\documentclass{article}

\usepackage{microtype}
\usepackage{graphicx}
\usepackage{subfigure}
\usepackage{booktabs} 

\usepackage{hyperref}

\usepackage[accepted]{icml2019}

\usepackage{graphicx}
\usepackage{amssymb}
\usepackage{amsfonts}
\usepackage{amsmath}
\usepackage{booktabs}
\usepackage{algorithm}
\usepackage{algorithmic}
\usepackage{wrapfig}
\usepackage{bm}
\usepackage{bbm}
\usepackage{breqn}
\usepackage{natbib}

\usepackage{todonotes}

\DeclareMathOperator*{\argmax}{arg\,max}

\icmltitlerunning{Learning Human Objectives by Evaluating Hypothetical Behavior}

\begin{document}

\twocolumn[
\icmltitle{Learning Human Objectives by Evaluating Hypothetical Behavior}



\icmlsetsymbol{equal}{*}

\begin{icmlauthorlist}
\icmlauthor{Siddharth Reddy}{berk,equal}
\icmlauthor{Anca D. Dragan}{berk}
\icmlauthor{Sergey Levine}{berk}
\icmlauthor{Shane Legg}{dm}
\icmlauthor{Jan Leike}{dm}
\end{icmlauthorlist}

\icmlaffiliation{berk}{University of California, Berkeley}
\icmlaffiliation{dm}{DeepMind}

\icmlcorrespondingauthor{Siddharth Reddy}{sgr@berkeley.edu}
\icmlcorrespondingauthor{Jan Leike}{jan@leike.name}

\icmlkeywords{Reinforcement Learning, Reward Modeling, Active Learning}

\vskip 0.3in
]



\printAffiliationsAndNotice{}  

\begin{abstract}
We seek to align agent behavior with a user's objectives in a reinforcement learning setting with unknown dynamics, an unknown reward function, and unknown unsafe states.
The user knows the rewards and unsafe states, but querying the user is expensive.
To address this challenge, we propose an algorithm that safely and interactively learns a model of the user's reward function.
We start with a generative model of initial states and a forward dynamics model trained on off-policy data.
Our method uses these models to synthesize hypothetical behaviors, asks the user to label the behaviors with rewards, and trains a neural network to predict the rewards.
The key idea is to actively synthesize the hypothetical behaviors from scratch by maximizing tractable proxies for the value of information, without interacting with the environment.
We call this method \emph{reward query synthesis via trajectory optimization}~(\mbox{ReQueST}).
We evaluate ReQueST with simulated users on a state-based 2D navigation task and the image-based Car Racing video game.
The results show that ReQueST significantly outperforms prior methods in learning reward models that transfer to new environments with different initial state distributions.
Moreover, ReQueST safely trains the reward model to detect unsafe states, and corrects reward hacking before deploying the agent.
\end{abstract}

\section{Introduction}

Users typically specify objectives for reinforcement learning~(RL) agents
through scalar-valued reward functions~\citep{sutton2018reinforcement}.
While users can easily define reward functions for tasks like playing games of Go or StarCraft, users may struggle to describe practical tasks like driving cars or controlling robotic arms in terms of rewards~\citep{hadfield2017inverse}.
Understanding user objectives in these settings can be challenging -- not only for machines, but also for humans modeling each other and introspecting on themselves~\citep{premack1978does}.

\begin{figure}[t]
    \centering
    \includegraphics[width=0.85\linewidth]{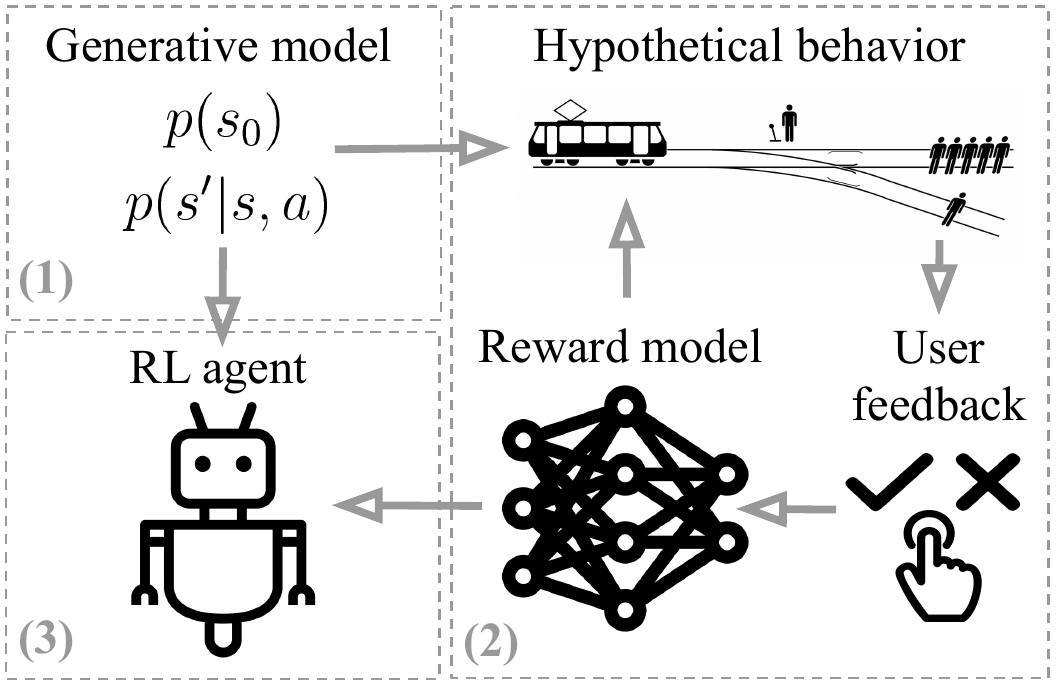}
    \caption{Our method learns a reward model from user feedback on hypothetical behaviors, then deploys a model-based reinforcement learning agent that optimizes the learned rewards.}
    \label{fig:schematic}
\end{figure}

For example, consider the \emph{trolley problem}~\citep{foot1967problem}: if you were the train conductor in Figure \ref{fig:schematic}, presented with the choice of either allowing multiple people to come to harm by letting the train continue on its current track, or harming one person by diverting the train, what would you do?
The answer depends on whether your value system leans toward consequentialism or deontological ethics -- a distinction that may not be captured by a reward function designed to evaluate common situations, in which ethical dilemmas like the trolley problem rarely occur.
In complex domains, the user may not be able to anticipate all possible agent behaviors and specify a reward function that accurately describes user preferences over those behaviors.

We address this problem by actively synthesizing hypothetical behaviors from scratch, and asking the user to label them with rewards.
Figure \ref{fig:schematic} describes our algorithm: using a generative model of initial states and a forward dynamics model trained on off-policy data, we synthesize hypothetical behaviors, ask the user to label the behaviors with rewards, and train a neural network to predict the rewards.
We repeat this process until the reward model converges, then deploy a model-based RL agent that optimizes the learned rewards.

\begin{figure}[t]
    \centering
    \includegraphics[width=\linewidth]{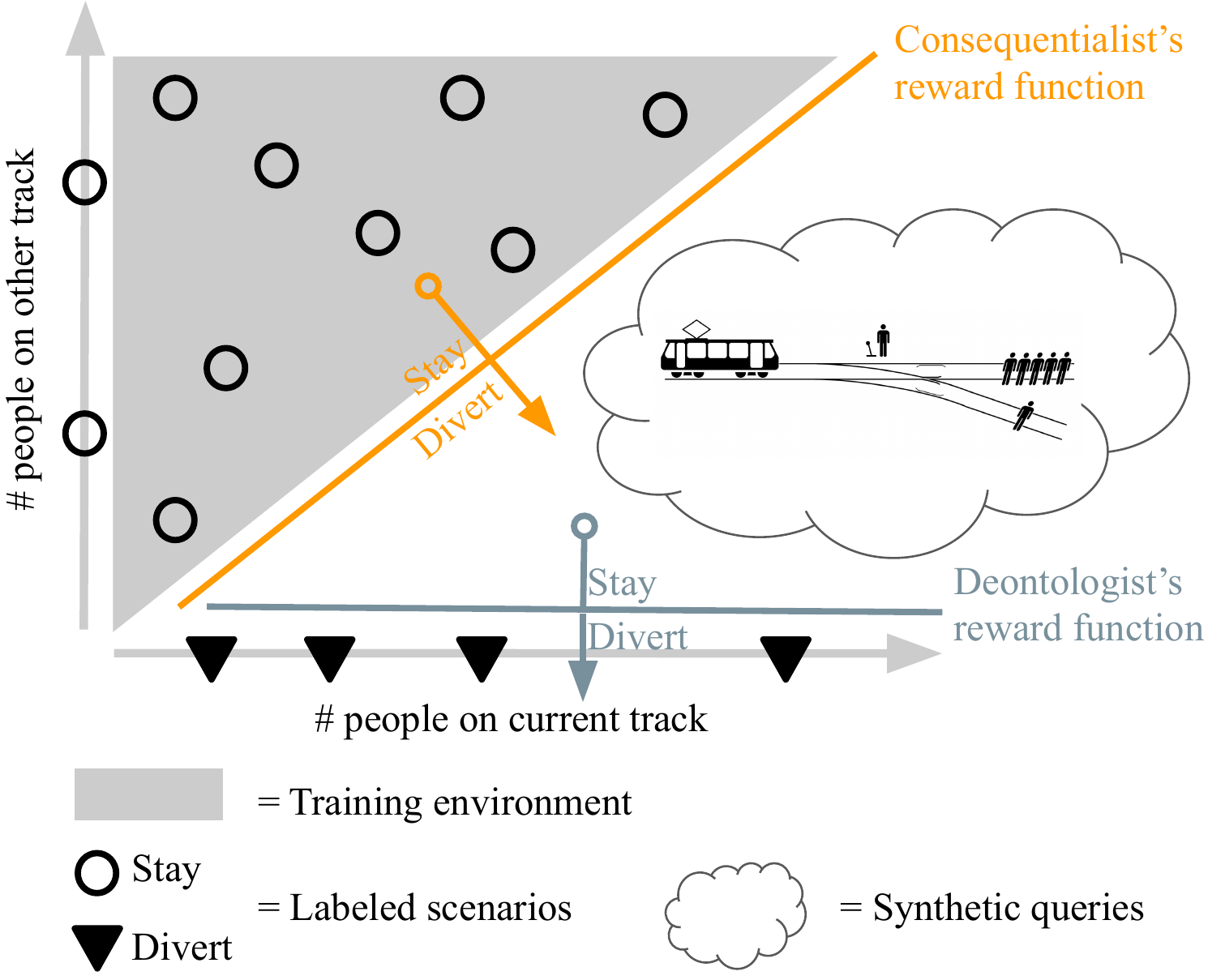}
    \caption{Our method automatically synthesizes hypotheticals like the trolley problem. Consider a training environment in which the following two states are common: either one of the tracks is empty, or there are fewer people on the current track than the other track. In these states, the consequentialist and deontologist reward functions agree. After asking the user to label these states, we are not able to determine which of the two is the true reward function, since both are consistent with the training data. Our method queries the user for labels at states where the value of information is highest: states where there are more people on the current track than the other track, but there are still some people on the other track. By eliciting user labels at these unlikely-but-informative states, we learn a reward model that more accurately captures the user's objectives.}
    \label{fig:trolley}
\end{figure}

The key idea in this paper is synthesizing informative hypotheticals (illustrated in Figure \ref{fig:trolley}).
Ideally, we would generate these hypotheticals by optimizing the value of information~(VOI; \citealp{savage1954foundations}), but the VOI is intractable for real-world domains with high-dimensional, continuous states.\footnote{The VOI is intractable because it requires computing an expectation over all possible trajectories, conditioned on the optimal policy for the updated reward model. See Section \ref{steptwoa} for details.} Instead, we use trajectory optimization to produce four types of hypotheticals that improve the reward model in different ways: behaviors that (1) maximize reward model uncertainty,\footnote{We measure uncertainty using the disagreement between an ensemble of reward models. See Section \ref{steptwoa} for details.} to elicit labels that are likely to change the updated reward model's outputs; (2) maximize predicted rewards, to detect and correct reward hacking; (3) minimize predicted rewards, to safely explore unsafe states; or (4) maximize novelty of trajectories regardless of predicted rewards, to improve the diversity of the training data.
To ensure that the hypothetical trajectories remain comprehensible to the user and resemble realistic behaviors, we use a generative model of initial states and a forward dynamics model for regularization.
We call this method \emph{reward query synthesis via trajectory optimization}~(ReQueST).

Our primary contribution is ReQueST: an algorithm that synthesizes hypothetical behaviors in order to safely and efficiently train neural network reward models in environments with high-dimensional, continuous states.
We evaluate ReQueST with simulated users in three domains: MNIST classification~\citep{lecun1998mnist}, a state-based 2D navigation task, and the image-based Car Racing video game in the OpenAI Gym~\citep{brockman2016openai}.
Our experiments show that ReQueST learns robust reward models that transfer to new environments with different initial state distributions, achieving at least 2x better final performance than baselines adapted from prior work (e.g., see Figure \ref{fig:robustness}).
In the navigation task, ReQueST safely learns to classify 100\% of unsafe states as unsafe and deploys an agent that never visits unsafe states, while the baselines fail to learn about even one unsafe state and deploy agents with a failure rate of 75\%.

\section{Related Work}

In this work, we align agent behavior with a user's objectives by learning a model of the user's reward function and training the agent via RL~\citep{russell1998learning,knox2012learning,knox2015framing,warnell2018deep,leike2018scalable}.
The idea behind modeling the user's reward function -- as opposed to the user's policy~\citep{ross2011reduction}, value function~\citep{reddy2018shared}, or advantage function~\citep{macglashan2017interactive} -- is to acquire a compact, transferable representation of the user's objectives; not just in the training environment, but also in new environments with different dynamics or initial states.

The closest prior work is on active learning methods for learning rewards from pairwise comparisons~\citep{dorsa2017active,biyik2018batch,wirth2017survey}, critiques~\citep{cui2018active}, demonstrations~\citep{ibarz2018reward,brown2019risk}, designs~\citep{mindermann2018active}, and numerical feedback~\citep{daniel2014active}.
ReQueST differs in three key ways: it produces query trajectories using a generative model, in a way that enables trading off between producing realistic vs. informative queries; it optimizes queries not only to reduce model uncertainty, but also to detect reward hacking and safely explore unsafe states; and it scales to learning neural network reward models that operate on high-dimensional, continuous state spaces.

ReQueST shares ideas with prior work~\citep{saunders2018trial,prakash2019improving} on learning to detect unsafe behaviors by initially seeking out catastrophes, selectively querying the user, and using model-based RL.
ReQueST differs primarily in that it learns a complete task specification, not just an unsafe state detector.
ReQueST is also complementary to prior work on safe exploration, which typically assumes a known reward function and side constraints, and focuses on ensuring that the agent never visits unsafe states during policy optimization~\citep{dalal2018safe,garcia2015comprehensive}.

\section{Learning Rewards from User Feedback on Hypothetical Behavior} \label{methods}

We formulate the reward modeling problem as follows.
We assume access to a training environment that follows a Markov Decision Process~(MDP; \citealp{sutton2018reinforcement}) with unknown state transition dynamics $\mathcal{T}$, unknown initial state distribution $\mathcal{S}_0^{\text{train}}$, and an unknown reward function $R$ that can be evaluated on specific inputs by querying the user.
We learn a model of the reward function $\hat{R}$ by querying the user for reward signals.
At test time, we train an RL agent with the learned reward function $\hat{R}$ in a new environment with the same dynamics $\mathcal{T}$, but a potentially different initial state distribution $\mathcal{S}_0^{\text{test}}$.
The goal is for the agent to perform well in the test environment with respect to the true reward function $R$.

Our approach to this problem is outlined in Figure \ref{fig:schematic}, and can be split into three steps.
In step (1) we use off-policy data to train a generative model $p_{\bm{\phi}}(\tau)$ that can be used to evaluate the likelihood of a trajectory $\tau = (s_0, a_0, s_1, a_1, ..., s_T)$.
This model enables us to synthesize hypothetical trajectories that can be shown to the user.
In step (2) we produce synthetic trajectories, which consist of sequences of state transitions $(s, a, s')$, that seek out different kinds of hypotheticals.
We ask the user to label each transition with a scalar reward $R(s, a, s')$, and fit a reward model $\hat{R}(s, a, s')$ using standard supervised learning techniques.\footnote{In principle, other methods, such as pairwise comparisons~\citep{wirth2017survey} or implicit feedback~\citep{cui2020empathic,xu2020accelerating}, could also be used to label the synthetic trajectories with user rewards.} In step (3) we use standard RL methods to train an agent using the learned rewards.
Since we typically learn a forward dynamics model as part of the generative model in step (1), we find that model-based RL is a good fit for training the agent in step (3).

\subsection{Learning a Generative Model of Trajectories} \label{stepone}

In order to synthesize hypothetical outcomes that may be unlikely to occur in the training environment, we cannot simply take actions in the training environment and collect the resulting trajectories, as is done in prior work.
Instead, we resort to training a generative model of trajectories, so that we can more efficiently sample unusual behaviors using the model.

In step (1) we collect off-policy data by interacting with the training environment in an unsupervised fashion; i.e., without the user in the loop.
To simplify our experiments, we sample trajectories $\tau$ by following random policies that explore a wide variety of states.\footnote{In principle, safe expert demonstrations could be used instead of random trajectories. We used random trajectories to simplify our experiments. See Section \ref{safety} for further discussion.} We use the observed trajectories to train a likelihood model,
\begin{equation} \label{dyn}
p_{\bm{\phi}}(\tau) \propto p_{\bm{\phi}}(s_0) \prod_{t=0}^{T-1} p_{\bm{\phi}}(s_{t+1} | s_t, a_t),
\end{equation}
where $p_{\bm{\phi}}(s_0)$ models the initial state distribution, $p_{\bm{\phi}}(s_{t+1} | s_t, a_t)$ models the forward dynamics, and $\bm{\phi}$ are the model parameters (e.g., neural network weights).
We train the model $p_{\bm{\phi}}$ using maximum-likelihood estimation, given the sampled trajectories.
As described in the next section, the likelihood model is helpful for regularizing the synthetic trajectories shown to the user.

In environments with high-dimensional, continuous states, such as images, we also train a state encoder $f_{\bm{\phi}} : \mathcal{S} \to \mathcal{Z}$ and decoder $f^{-1}_{\bm{\phi}} : \mathcal{Z} \to \mathcal{S}$, where $\mathcal{S} = \mathbb{R}^n$, $\mathcal{Z} = \mathbb{R}^d$, and $d << n$.
As described in Section \ref{steptwob}, embedding states in a low-dimensional latent space $\mathcal{Z}$ is helpful for trajectory optimization.
In our experiments, we train $f_{\bm{\phi}}$ and $f^{-1}_{\bm{\phi}}$ using the variational auto-encoder method~(VAE; \citealp{kingma2013auto}).

\subsection{Representing the Reward Model as a Classifier} \label{rewasclf}

Our goal is to learn a model $\hat{R}$ of the user's reward function.
In step (2) we represent $\hat{R}$ by classifying state transitions as good, unsafe, or neutral -- similar to \citet{cui2018active} -- and assigning a known, constant reward to each of these three categories:
\begin{equation} \label{eq:rew-clf}
\hat{R}(s, a, s') = \sum_{c \in \{\text{good}, \text{unsafe}, \text{neutral}\}} p_{\bm{\theta}}(c | s, a, s') R_c,
\end{equation}
where $p_{\bm{\theta}}(c | s, a, s') = \frac{1}{m}\sum_{i=1}^m p_{\bm{\theta}_i}(c | s, a, s')$ is the mean of an ensemble of $m$ classifiers $\{p_{\bm{\theta}_i}\}_{i=1}^m$, and $\bm{\theta}_i$ are the weights of the $i$-th neural network in the ensemble.
$R_c$ is the constant reward for any state transition in class $c$, where $R_{\text{unsafe}} \leq R_{\text{neutral}} \leq R_{\text{good}}$.
Modeling the reward function as a classifier simplifies our experiments and makes it easier for the user to provide labels.
In principle, our method can also work with other architectures, such as a more straightforward regression model $\hat{R} = R_{\bm{\theta}}$.
As described in Section \ref{steptwoa}, we use an ensemble method to model uncertainty.

\subsection{Designing Objectives for Informative Queries} \label{steptwoa}

Our approach to reward modeling involves asking the user to label trajectories with reward signals.
In step (2) we synthesize query trajectories to elicit user labels that are informative for learning the reward model.

To generate a useful query, we synthesize a trajectory $\tau$ that maximizes an acquisition function~(AF) denoted by $J(\tau)$.
The AF evaluates how useful it would be to elicit reward labels for $\tau$, then update the reward model given the newly-labeled data.
Since we do not assume knowledge of the test environment where the agent is deployed, we cannot optimize the ideal AF: the value of information~(VOI; \citealp{savage1954foundations}), defined as the gain in performance of an agent that optimizes the updated reward model in the test environment.
Prior work on active learning tackles this problem by optimizing proxies for VOI~\citep{settles2009active}.
We use AFs adapted from prior work, as well as novel AFs that are particularly useful for reward modeling.

In this work, we use four AFs that are easy to optimize for neural network reward models.
The first AF $J_u(\tau)$ maximizes reward model uncertainty, eliciting user labels for behaviors that are likely to change the updated reward model's outputs.
$J_u$ is a proxy for VOI, since improving the agent requires improving the predicted rewards.
The second AF $J_+(\tau)$ maximizes predicted rewards, surfacing behaviors for which the reward model might be incorrectly predicting high rewards.
$J_+$ is another useful heuristic for reward modeling, since preventing reward hacking improves agent performance.
The third AF $J_-(\tau)$ minimizes predicted rewards, adding unsafe behaviors to the training data.
While we do not consider $J_-$ to be a proxy for VOI, we find it helpful empirically for training neural network reward models, since it helps to balance the number of unsafe states (vs. neutral or good states) in the training data.
The fourth AF $J_n(\tau)$ maximizes the novelty of training data, encouraging uniform coverage of the space of behaviors regardless of their predicted reward.
$J_n$ is a na\"ive proxy for VOI, but tends to be helpful in practice due to the difficulty of estimating the uncertainty of neural network reward models for $J_u$.

\noindent\textbf{Maximizing uncertainty.}
The first AF $J_u(\tau)$ implements one of the simplest query selection strategies from the active learning literature: uncertainty sampling~\citep{lewis1994sequential}.
The idea is to elicit labels for examples that the model is least certain how to label, and thus reduce model uncertainty.
To do so, we train an ensemble of neural network reward models, and generate trajectories that maximize the disagreement between ensemble members.
Following \citet{lakshminarayanan2017simple}, we measure ensemble disagreement using the average KL-divergence between the output of a single ensemble member and the ensemble mean,
\begin{align} \label{eq:disag}
J_u(\tau) = \frac{1}{|\tau|} \sum_{(s, a, s') \in \tau} \frac{1}{m} \sum_{i=1}^m D_{\mathrm{KL}}(& p_{\bm{\theta}_i}(c | s, a, s') \nonumber \\
& \|\; p_{\bm{\theta}}(c | s, a, s')),
\end{align}
where $p_{\bm{\theta}}$ is the reward classifier defined in Section \ref{rewasclf}.
Although more sophisticated methods of modeling uncertainty in neural networks exist~\citep{gal2016uncertainty}, we find that ensemble disagreement works well in practice.\footnote{We did not compare to other ensemble-based approximations, such as mutual information~\citep{houlsby2011bayesian}.}

\noindent\textbf{Maximizing reward.}
The second AF $J_+(\tau)$ is intended to detect examples of false positives, or `reward hacking': behaviors for which the reward model incorrectly outputs high reward~\citep{amodei2016concrete,christiano2017deep}.
The idea is to show the user what the reward model predicts to be good behavior, with the expectation that some of these behaviors are actually suboptimal, and will be labeled as such by the user.
To do so, we simply synthesize trajectories that maximize $J_+(\tau) = \sum_{(s, a, s') \in \tau} \hat{R}(s, a, s')$.

\noindent\textbf{Minimizing reward.}
The third AF $J_-(\tau)$ is intended to augment the training data with more examples of unsafe states than would normally be encountered, e.g., by a reward-maximizing agent acting in the training environment.
The idea is to show the user what the reward model considers to be unsafe behavior, with the expectation that the past training data may not contain egregiously unsafe behaviors, and that it would be helpful for the user to confirm whether the model has captured the correct notion of unsafe states.
To do so, we produce trajectories that maximize $J_-(\tau) = -J_+(\tau)$.

\noindent\textbf{Maximizing novelty.}
The fourth AF $J_n(\tau)$ is intended to produce novel trajectories that differ from those already in the training data, regardless of their predicted reward; akin to prior work on geometric AFs~\citep{sener2017geometric}.
This is especially helpful early during training, when uncertainty estimates are not accurate, and the reward model has not yet captured interesting notions of reward-maximizing and reward-minimizing behavior.
To do so, we produce trajectories $\tau$ that maximize the distance between $\tau$ and previously-labeled trajectories $\tau' \in \mathcal{D}$,
\begin{equation} \label{eq:nov}
J_n(\tau) = \frac{1}{|\mathcal{D}|}\sum_{\tau' \in \mathcal{D}} d(\tau, \tau').
\end{equation}
In this work, we use a distance function that computes the Euclidean distance between state embeddings,
\begin{equation} \label{eq:dist}
d(\tau, \tau') = \frac{1}{|\tau| |\tau'|}\sum_{s \in \tau, s' \in \tau'} -e^{-\|f_{\bm{\phi}}(s_t) - f_{\bm{\phi}}(s'_t)\|_2},
\end{equation}
where $f_{\bm{\phi}}$ is the state encoder trained in step (1).

For the sake of simplicity, we synthesize a separate trajectory for each of the AFs $J_u$, $J_+$, $J_-$, and $J_n$.
In principle, multiple AFs could be combined to form hybrid AFs.
For example, optimizing $J'(\tau) = J_+(\tau) + J_n(\tau)$ could yield trajectories that simultaneously maximize rewards and novelty.

\subsection{Query Synthesis via Trajectory Optimization} \label{steptwob}

We synthesize a query trajectory $\tau_{\text{query}}$ by solving the optimization problem,
\begin{equation} \label{eq:query-synth}
\tau_{\text{query}} = \max_{z_0, a_0, z_1, ..., z_T} J(\tau) + \lambda \log{p_{\bm{\phi}}(\tau)},
\end{equation}
where $z_t$ is the embedding of state $s_t$ in the latent space of the encoder $f$ trained in step (1), $\tau = (f^{-1}(z_0), a_0, f^{-1}(z_1), a_1, ..., f^{-1}(z_T))$ is the decoded trajectory, $J$ is the acquisition function (Section \ref{steptwoa}), $\lambda \in \mathbb{R}_{\geq 0}$ is a regularization constant, and $p_{\bm{\phi}}$ is the generative model of trajectories (Section \ref{stepone}).
In this work, we assume $p_{\bm{\phi}}(\tau)$ is differentiable, and optimize $\tau_{\text{query}}$ using Adam~\citep{kingma2014adam}.\footnote{Our method can be extended to settings where $p_{\bm{\phi}}(\tau)$ is not differentiable, by using a gradient-free optimization method to synthesize $\tau_{\text{query}}$. This can be helpful, e.g., when using a non-differentiable simulator to model the environment.} Optimizing low-dimensional, latent states $z$ instead of high-dimensional, raw states $s$ reduces computational requirements, and regularizes the optimized states to be more realistic.\footnote{Our approach to query synthesis draws inspiration from direct collocation methods in the trajectory optimization literature~\citep{betts2010practical}, feature visualization methods in the neural network interpretability literature~\citep{olah2017feature}, and prior work on active learning with deep generative models~\citep{huijser2017active}.}

The regularization constant $\lambda$ from \autoref{eq:query-synth} controls the trade-off between how realistic $\tau_{\text{query}}$ is and how aggressively it maximizes the AF.
Setting $\lambda = 0$ can result in query trajectories that are incomprehensible to the user and unlikely to be seen in the test environment, while setting $\lambda$ to a high value can constrain the query trajectories from seeking interesting hypotheticals.
The experiments in Section \ref{sweep} analyze this trade-off in further detail.

\begin{algorithm}[t]
\small
\begin{algorithmic}[1]
\STATE Require $\lambda, p_{\bm{\phi}}$
\STATE{Initialize $\mathcal{D} \leftarrow \emptyset$ \label{lst:line:init-data}}
\WHILE {$\bm{\theta}$ not converged}
  \FOR {$J \in \{J_u, J_+, J_-, J_n\}$ \label{lst:line:acqs}}
  \STATE{$\tau_{\text{query}} \leftarrow \max_{\tau} J(\tau) + \lambda \log{p_{\bm{\phi}}(\tau)}$ \label{lst:line:qsynth}}
  \FOR {$(s, a, s') \in \tau_{\text{query}}$}
      \STATE{$c \leftarrow c \sim p_{\text{user}}(c | s, a, s')$ \COMMENT{Query the user} \label{lst:line:user-query}}
      \STATE{$\mathcal{D} \leftarrow \mathcal{D} \cup \{(s, a, s', c)\}$}
  \ENDFOR
  \ENDFOR
  \FOR {$i \in \{1, 2, ..., m\}$}
  \STATE{$\bm{\theta}_i \leftarrow \argmax_{\bm{\theta}_i} \sum_{(s, a, s', c) \in \mathcal{D}} \log{p_{\bm{\theta}_i}(c | s, a, s')}$ \label{lst:line:mle}}
  \ENDFOR
\ENDWHILE
\STATE{Return reward model $\hat{R}$ \COMMENT{Defined via $\bm{\theta}$ in Equation \ref{eq:rew-clf}}}
\end{algorithmic}
\caption{Reward Query Synthesis\\via Trajectory Optimization~(ReQueST)}
\label{alg:rqst-alg}
\end{algorithm}

Our reward modeling algorithm is summarized in Algorithm \ref{alg:rqst-alg}.
Given a generative model of trajectories $p_{\bm{\phi}}(\tau)$, it generates one query trajectory $\tau_{\text{query}}$ for each of the four AFs, asks the user to label the states in the query trajectories, retrains the reward model ensemble $\{\bm{\theta}_i\}_{i=1}^m$ on the updated training data $\mathcal{D}$ using maximum-likelihood estimation,\footnote{Note that, in line 12 of Algorithm \ref{alg:rqst-alg}, we train each ensemble member on all of the data $\mathcal{D}$, instead of a random subset of the data (i.e., bootstrapping). As in \citet{lakshminarayanan2017simple}, we find that simply training each reward network $\bm{\theta}_i$ using a different random seed works well in practice for modeling uncertainty.} and repeats this process until the user is satisfied with the outputs of the reward model.
The ablation study in Section \ref{ablation} analyzes the effect of using different subsets of the four AFs to generate queries.

\subsection{Deploying a Model-Based RL Agent} \label{stepthree}

Given the learned reward model $\hat{R}$, the agent can, in principle, be trained using any RL algorithm in step (3).
In practice, since our method learns a forward dynamics model in step (1), we find that model-based RL is a good fit for training the agent in step (3).
In this work, we deploy an agent $\pi_{\text{mpc}}$ that combines planning with model-predictive control~(MPC):
\begin{align} \label{eq:mpc}
\pi_{\text{mpc}}(a | s) = & \mathbbm{1}\left[ a = \argmax_{a_0} \max_{a_1, \ldots, a_H} \hat{R}(s, a_0) \right. \nonumber \\
& + \hat{R}(\mathbb{E}_{\bm{\phi}}[s_1 | s, a_0], a_1) + ... \nonumber \\
& \left. + \hat{R}(\mathbb{E}_{\bm{\phi}}[s_H | s, a_0, a_1, ..., a_{H-1}], a_H) \right],
\end{align}
where the future states $\mathbb{E}_{\bm{\phi}}[s_t | s, a_0, a_1, ..., a_{t-1}]$ are predicted using the forward dynamics model $p_{\bm{\phi}}$ trained in step (1), $H$ is the planning horizon, and $\hat{R}$ is the reward model trained in step (2).
We solve the optimization problem in the right-hand side using Adam~\citep{kingma2014adam}.

\subsection{Safe Exploration} \label{safety}

One of the benefits of our method is that, since it learns from synthetic trajectories instead of real trajectories, it only has to imagine visiting unsafe states, instead of actually visiting them.
Although unsafe states may be visited during unsupervised exploration of the environment for training the generative model in step (1), the same generative model can be reused to learn reward models for any number of future tasks.
Hence, the cost of visiting a fixed number of unsafe states in step (1) can be amortized across a large number of tasks in step (2).
We could also train the generative model on other types of off-policy data instead, including safe expert demonstrations and examples of past failures.

Another benefit of our method is that, as part of the data collection process in step (2), the user gets to observe query trajectories that reveal what the reward model has learned.
Thus, the user can choose to stop providing feedback when they are satisfied with the reward model's notions of reward-maximizing and reward-minimizing behaviors; and when they see that uncertainty-maximizing queries are genuinely ambiguous, instead of merely uncertain to the model while being easy for the user to judge.
This provides a safer alternative to debugging a reward model by immediately deploying the agent and observing its behavior without directly inspecting the reward model beforehand.

\section{Experimental Evaluation} \label{exp-overview}

We seek to answer the following questions. \textbf{Q1}: Does synthesizing hypothetical trajectories elicit more informative labels than rolling out a policy in the training environment? \textbf{Q2}: Can our method detect and correct reward hacking? \textbf{Q3}: Can our method safely learn about unsafe states? \textbf{Q4}: Do the proposed AFs improve upon random sampling from the generative model? \textbf{Q5}: How does the regularization constant $\lambda$ control the trade-off between realistic and informative queries? \textbf{Q6}: How much do each of the four AFs contribute to performance?

\begin{figure}[t]
    \centering
    \includegraphics[width=0.49\linewidth]{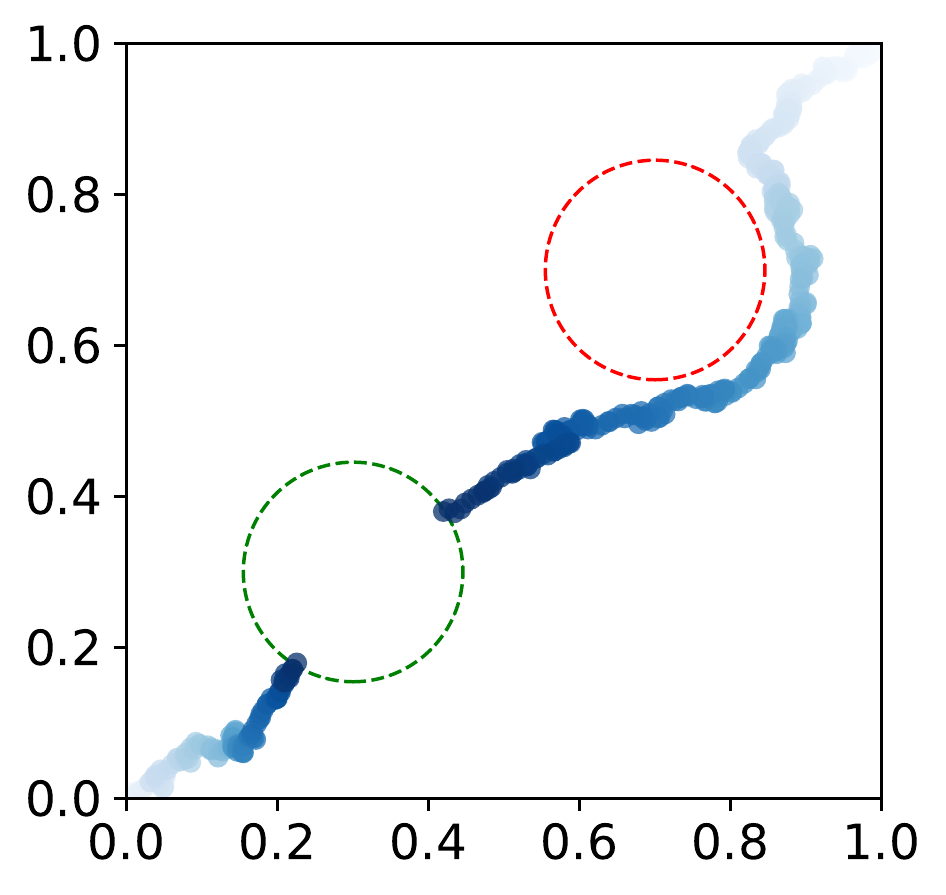}
    \includegraphics[width=0.49\linewidth]{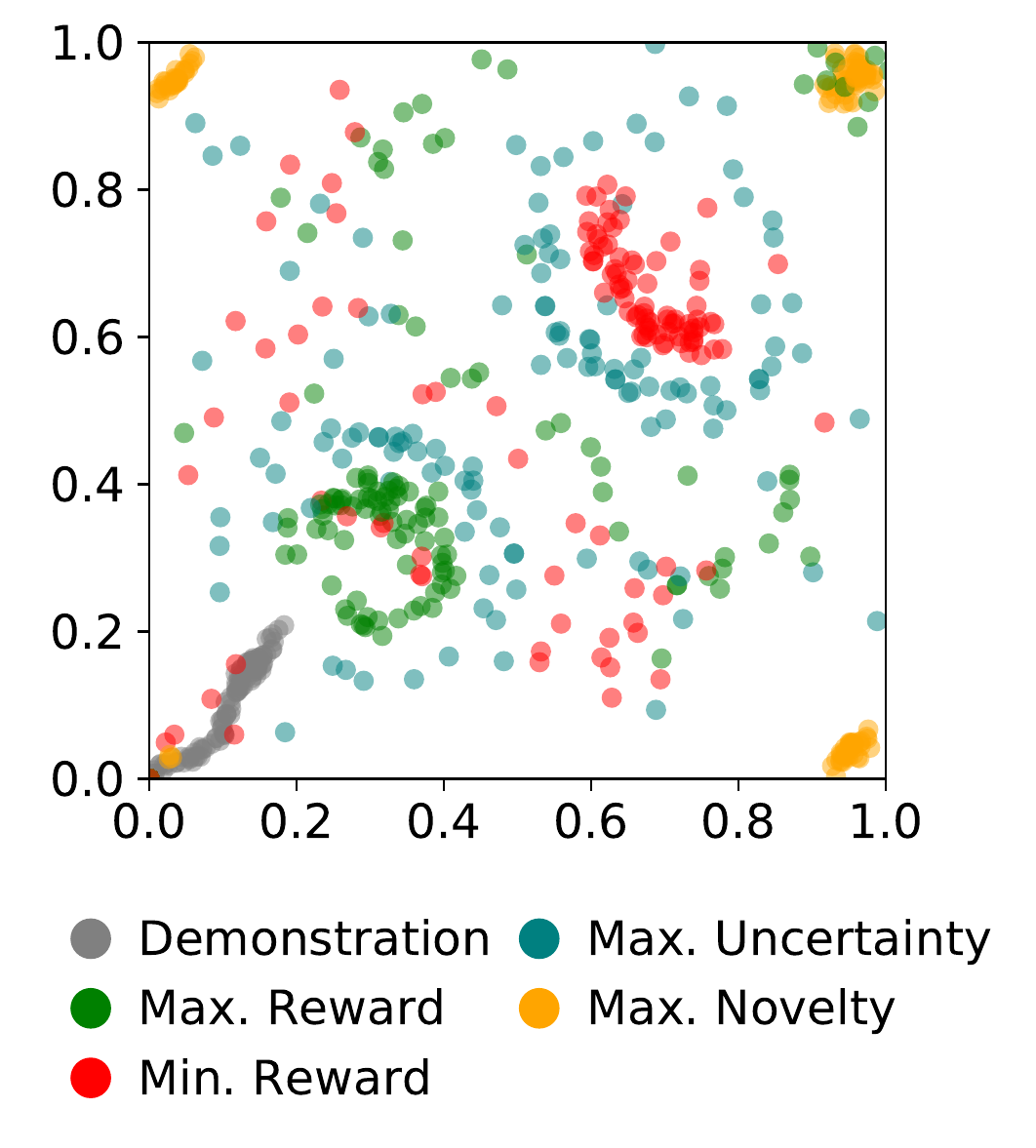}
    \caption{Left: The 2D navigation task, where the agent navigates to the goal region (green) in the lower left while avoiding the trap region (red) in the upper right. The agent starts in the lower left corner in the training environment, and starts in the upper right corner in the test environment. Right: Examples of hypothetical states synthesized throughout learning, illustrating the qualitative differences in the behaviors targeted by each AF.}
    \label{fig:pointmass}
\end{figure}

To answer these questions under ideal assumptions, we run experiments in three domains -- MNIST~\citep{lecun1998mnist}, state-based 2D navigation (Figure \ref{fig:pointmass}), and image-based Car Racing from the OpenAI Gym~\citep{brockman2016openai} -- with simulated users that label trajectories using a ground-truth reward function.
In each domain, we setup a training environment with initial state distribution $\mathcal{S}_0^{\text{train}}$, and a test environment with initial state distribution $\mathcal{S}_0^{\text{test}}$, as described in Section \ref{methods}.
In many real-world settings, the user can help initialize the reward model by providing a small number of (suboptimal) demonstrations and labeling them with rewards.
Hence, we initialize the training data $\mathcal{D}$ in line 2 of Algorithm \ref{alg:rqst-alg} with a small set of labeled, suboptimal, user demonstrations collected in the training environment.

\noindent\textbf{MNIST classification.}
This domain enables us to focus on testing the active learning component of our method, since the standard digit classification task does not involve sequential decision-making.
Here, the initial state $s_0 \in \mathbb{R}^{28 \times 28}$ is a grayscale image of a handwritten digit, and the action $a \in \{0, 1, ..., 9\}$ is a discrete classification.
When we generate queries, we synthesize an image $s_0$, and ask the simulated user to label it with an action $a$.
The initial state distribution of the training environment $\mathcal{S}_0^{\text{train}}$ puts a probability of $1$ on sampling $s_0 \in \{5, 6, 7, 8, 9\}$, and a probability of $0$ on sampling $s_0 \in \{0, 1, 2, 3, 4\}$.
We intentionally introduce a significant shift in the state distribution between the training and test environments, by setting the initial state distribution of the test environment $\mathcal{S}_0^{\text{test}}$ to the complement of $\mathcal{S}_0^{\text{train}}$; i.e., putting a probability of $1$ on sampling $s_0 \in \{0, 1, 2, 3, 4\}$, and a probability of $0$ on sampling $s_0 \in \{5, 6, 7, 8, 9\}$.
This mismatch is intended to test the robustness of the learned classifier; i.e., how well it performs under distribution shift.
We train a state encoder $f_{\bm{\phi}}$ and decoder $f^{-1}_{\bm{\phi}}$ in step (1) by training a VAE with an 8-dimensional latent space $Z = \mathbb{R}^8$ on all the images in the MNIST training set.\footnote{Note that this differs from the random sampling method for collecting off-policy data described in Section \ref{stepone}. Though the initial state distribution of the training environment is a uniform distribution over $\{5, 6, 7, 8, 9\}$, we train the generative model on all the digits $\{0, 1, 2, ..., 9\}$. This simplifies our experiments, and enables ReQueST to synthesize hypothetical digits from $\{0, 1, 2, 3, 4\}$.}

\noindent\textbf{State-based 2D navigation.}
This domain enables us to focus on the challenges of sequential decision-making, without dealing with high-dimensional states.
Here, the state $s \in \mathbb{R}^2$ is the agent's position, and the action $a \in \mathbb{R}^2$ is a velocity vector.
The task requires navigating to a target region, while avoiding a trap region.
The simulated user labels a state transition $(s, a, s')$ with category $c \in \{\text{good}, \text{unsafe}, \text{neutral}\}$, by looking at the state $s'$, and identifying whether it is inside the goal region (good), inside the trap region (unsafe), or outside both regions (neutral).
The initial state distribution of the training environment $\mathcal{S}_0^{\text{train}}$ is a delta function at the origin: $p(s_0) = \mathbbm{1}[s_0 = (0, 0)]$.
We intentionally introduce a significant shift in the state distribution between the training and test environments, by setting the initial state distribution of the test environment $\mathcal{S}_0^{\text{test}}$ to a delta function at the opposite corner of the unit square: $p(s_0) = \mathbbm{1}[s_0 = (1, 1)]$.
As in MNIST, this mismatch is intended to test the generalization of reward models.
The task is harder to complete in the test environment, since the agent starts closer to the trap, and must navigate around the trap to reach the goal (Figure \ref{fig:pointmass}).

\begin{figure*}[t]
    \centering
    \includegraphics[height=0.23\textheight]{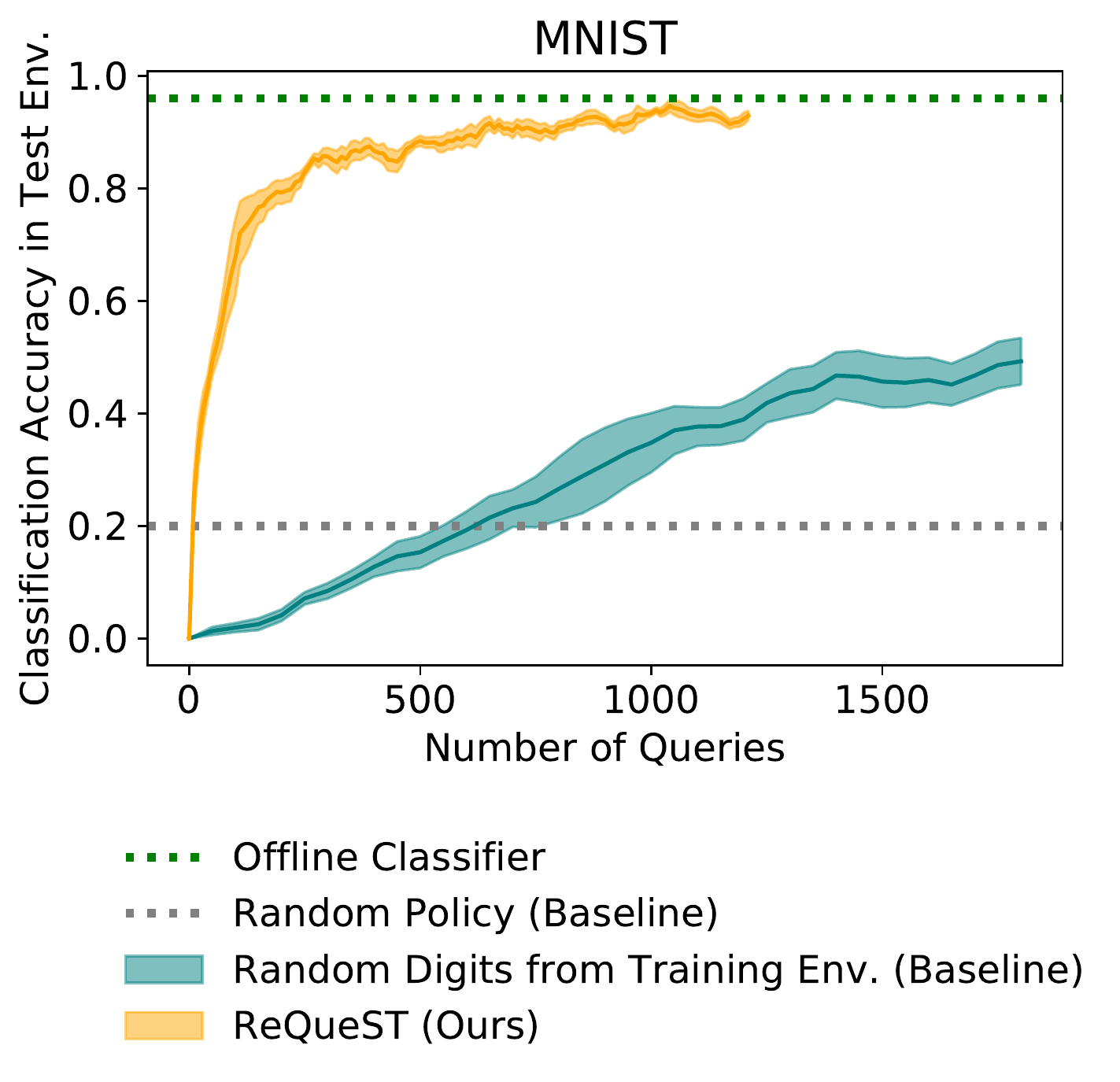}
    \includegraphics[height=0.23\textheight]{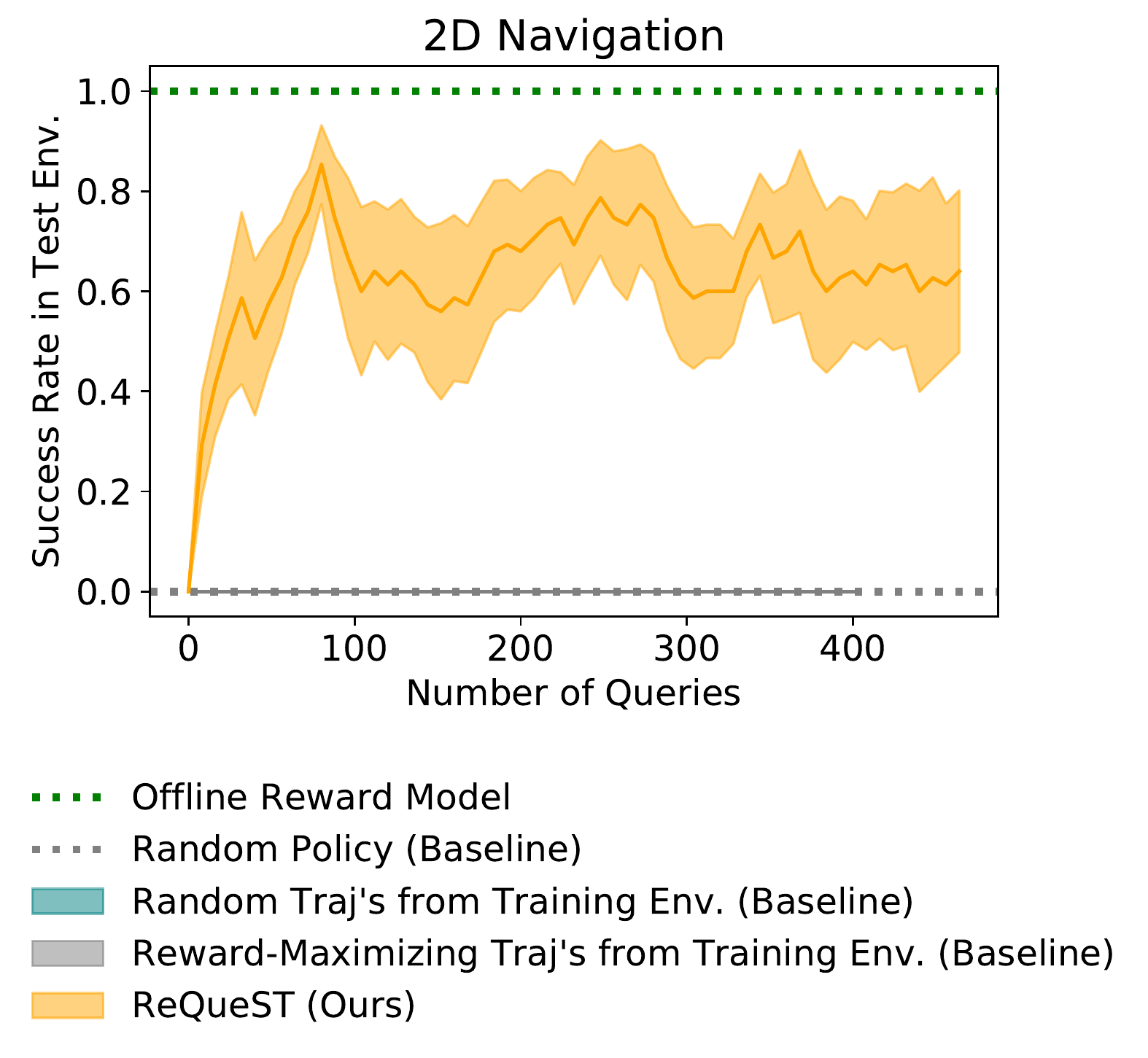}
    \includegraphics[height=0.23\textheight]{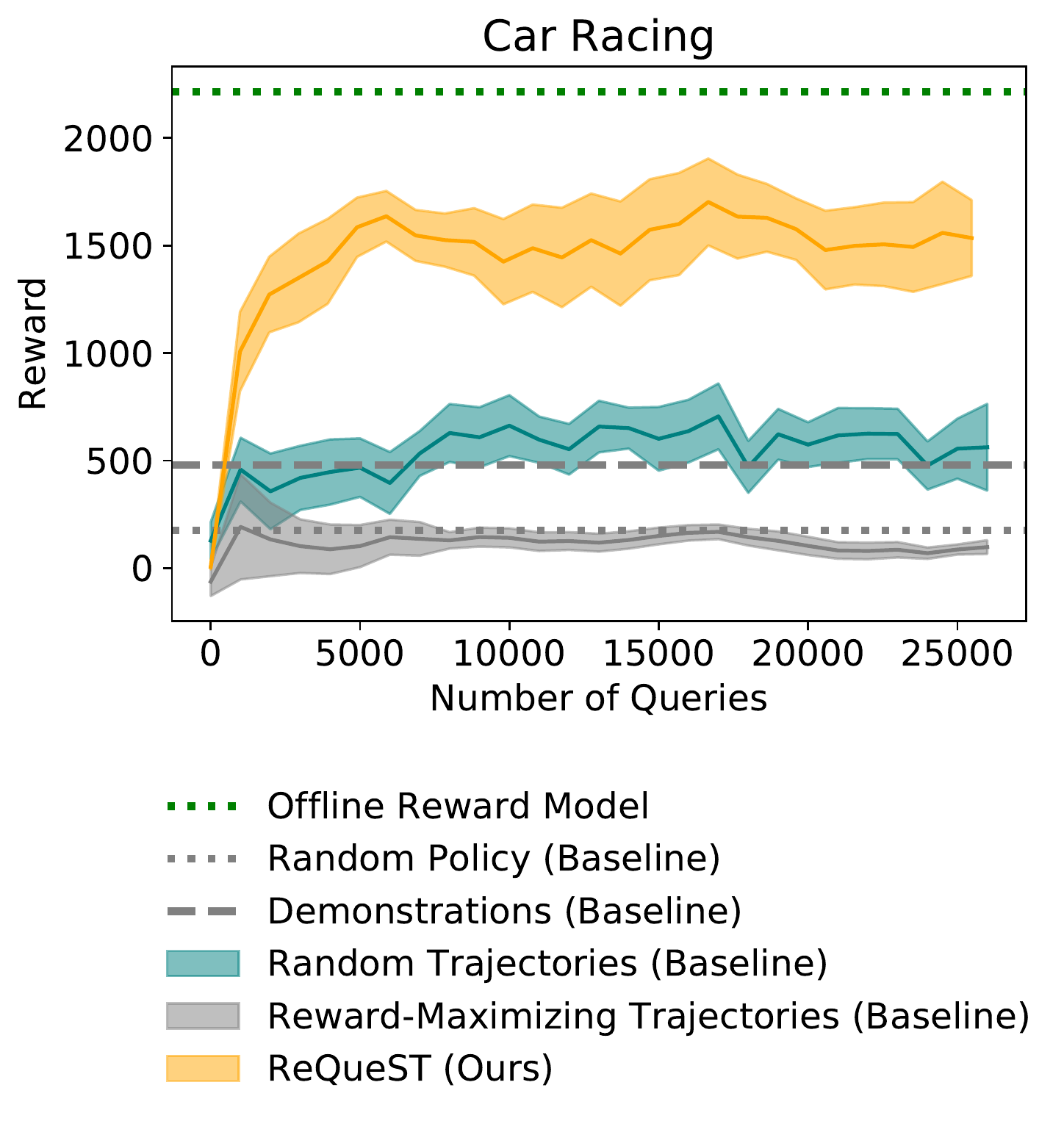}
    \caption{Experiments that address \textbf{Q1} -- does synthesizing hypothetical trajectories elicit more informative labels than rolling out a policy in the training environment? -- by comparing our method, which uses synthetic trajectories, to baselines that only use real trajectories generated in the training environment. The results on MNIST, 2D navigation, and Car Racing show that our method (orange) significantly outperforms the baselines (blue and gray), which never succeed in 2D navigation. The x-axis represents the number of queries to the user, where each query elicits a label for a single state transition $(s, a, s')$. The shaded areas show standard error over three random seeds.}
    \label{fig:robustness}
\end{figure*}

\noindent\textbf{Image-based Car Racing.}
This domain enables us to test whether our method scales to learning sequential tasks with high-dimensional states.
Here, the state $s \in \mathbb{R}^{64 \times 64 \times 3}$ is an RGB image with a top-down view of the car (Figure \ref{fig:carracing-screenshot} in the appendix), and the action $a \in \mathbb{R}^3$ controls steering, gas, and brake.
The simulated user labels a state transition $(s, a, s')$ with category $c \in \{\text{good}, \text{unsafe}, \text{neutral}\}$, by looking at the state $s'$, and identifying whether it shows the car driving onto a new patch of road (good), off-road (unsafe), or in a previously-visited road patch (neutral).
Here, we set the same initial state distribution for the training and test environments, since the reward modeling problem is challenging even when the initial state distributions are identical.
We train a generative model in step (1) using the unsupervised approach in \citet{ha2018recurrent}, which trains a VAE that compresses images, a recurrent dynamics model that predicts state transitions under partial observability, and a mixture density network that predicts stochastic transitions.

Section \ref{imp-deets} in the appendix discusses the setup of each domain, including the methods used to train the generative model in step (1), in further detail.

\subsection{Robustness Compared to Baselines} \label{exp-robustness}

Our first experiment tests whether our method can learn a reward model that is robust enough to perform well in the test environment, and tracks how many queries to the user it takes to learn an accurate reward model.

\noindent\textbf{Manipulated factors.}
To answer \textbf{Q1}, we compare our method to a baseline that, instead of generating hypothetical trajectories for the user to label, generates trajectories by rolling out a policy that optimizes the current reward model in the training environment -- an approach adapted from prior work~\citep{christiano2017deep}.
The baseline generates $\tau_{\text{query}}$ in line 5 of Algorithm \ref{alg:rqst-alg} by rolling out the MPC policy in Equation \ref{eq:mpc}, instead of solving the optimization problem in Equation \ref{eq:query-synth}.
To test how generating queries using a reward-maximizing policy compares to using a policy that does not depend on the reward model, we also evaluate a simpler baseline that generates query trajectories using a uniform random policy, instead of the MPC policy.

\noindent\textbf{Dependent measures.}
We measure performance in MNIST using the agent's classification accuracy in the test environment; in 2D navigation, the agent's success rate at reaching the goal while avoiding the trap in the test environment; and in Car Racing, the agent's true reward, which gives a bonus for driving onto new patches of road, and penalizes going off-road.\footnote{We also measure performance in the training environment, without state distribution shift. See Figure \ref{fig:trainenvres} in the appendix for details.} We establish a lower bound on performance using a uniform random policy, and an upper bound by deploying an MPC agent equipped with a reward model trained on a large, offline dataset of 100 expert trajectories and 100 random trajectories containing balanced classes of good, unsafe, and neutral state transitions.

\noindent\textbf{Analysis.}
The results in Figure \ref{fig:robustness} show that our method produces reward models that transfer to the test environment better than the baselines.
Our method also learns to outperform the suboptimal demonstrations used to initialize the reward model (Figure \ref{fig:superhuman} in the appendix).

In MNIST, our method performs substantially better than the baseline, which samples queries $s_0$ from the initial state distribution of the training environment.
The reason is simple: the initial state distribution of the test environment differs significantly from that of the training environment.
Since our method is not restricted to sampling from the training environment, it performs better than the baseline.

In 2D navigation, our method substantially outperforms both baselines, which never succeed in the test environment.
This is unsurprising, since the training environment is set up in such a way that, because the agent starts out in the lower left corner, they rarely visit the trap region in the upper right by simply taking actions -- whether reward-maximizing actions (as in the first baseline), or uniform random actions (as in the second baseline).
Hence, when a reward model trained by the baselines is transferred to the test environment, it is not aware of the trap, so the agent tends to get caught in the trap on its way to the goal.
Our method, however, is not restricted to feasible trajectories in the training environment, and can potentially query the label for any position in the environment -- including the trap~(see \autoref{fig:pointmass}).
Hence, our method learns a reward model that is aware of the trap, which enables the agent to navigate around it in the test environment.

In Car Racing, our method outperforms both baselines.
This is mostly due to the fact that the baselines tend to generate queries that are not diverse and rarely visit unsafe states, so the resulting reward models are not able to accurately distinguish between good, unsafe, and neutral states.
Our method, on the other hand, explicitly seeks out a wide variety of states by maximizing the four AFs, which leads to more diverse training data, and a more accurate reward model.

\subsection{Detecting Reward Hacking}

\begin{figure}[t]
    \centering
    \includegraphics[width=0.85\linewidth]{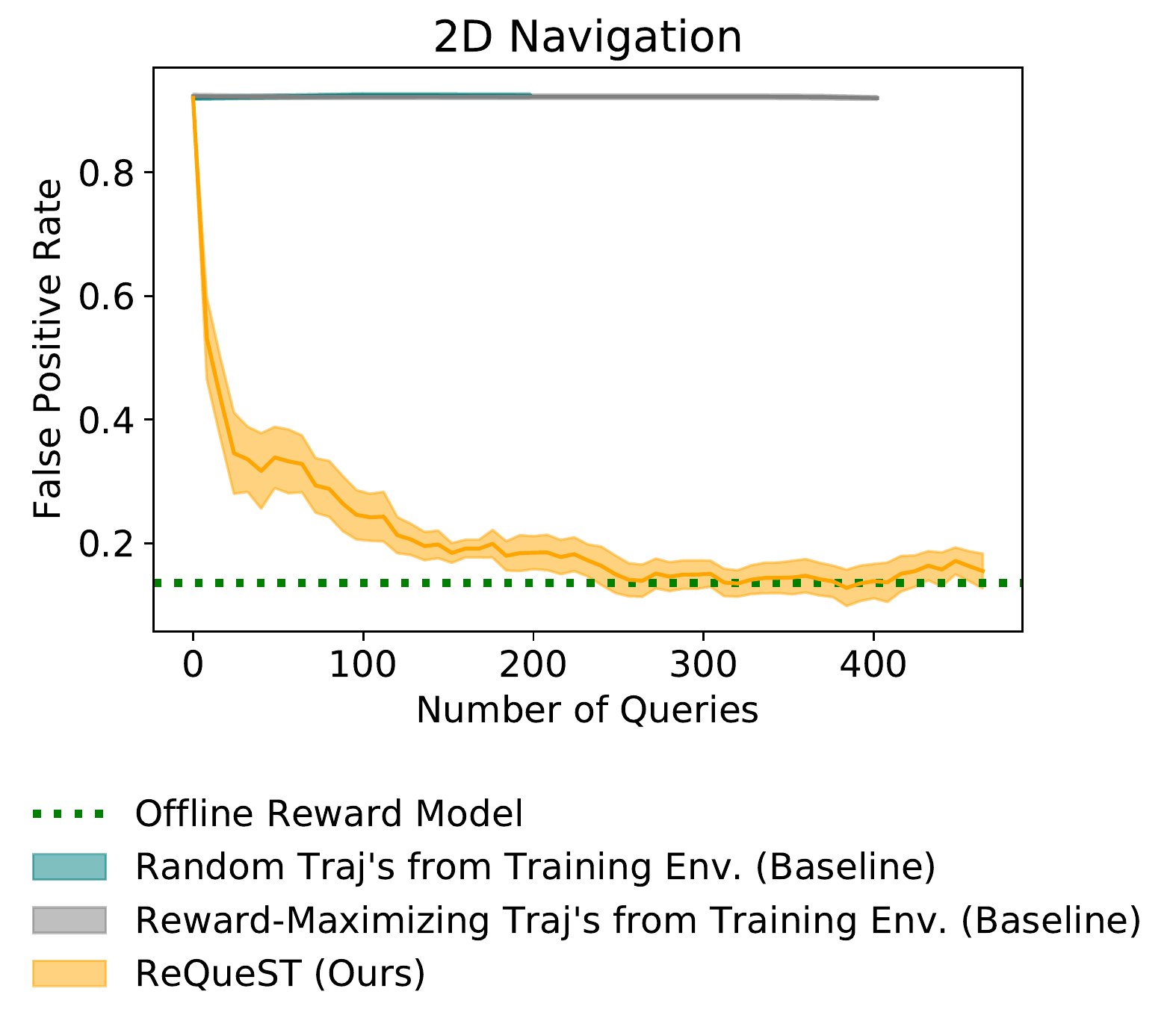}
    \caption{Experiments that address \textbf{Q2} -- can our method detect and correct reward hacking? -- by comparing our method, which uses synthetic trajectories, to baselines that only use real trajectories generated in the training environment. The results on 2D navigation show that our method (orange) significantly outperforms the baselines (blue and gray). The x-axis represents the number of queries to the user, where each query elicits a label for a single state transition $(s, a, s')$. The shaded areas show standard error over three random seeds.}
    \label{fig:rewhack}
\end{figure}

One of the benefits of our method is that it can detect and correct reward hacking before deploying the agent, using reward-maximizing synthetic queries.
In the next experiment, we test this claim.

\noindent\textbf{Manipulated factors.}
We replicate the experimental setup in Section \ref{exp-robustness} for 2D navigation, including the same baselines.

\noindent\textbf{Dependent measures.}
We measure performance using the false positive rate of the reward model: the fraction of neutral or unsafe states incorrectly classified as good, evaluated on the offline dataset of trajectories described in Section \ref{exp-robustness}.
A reward model that outputs false positives is susceptible to reward hacking, since a reward-maximizing agent can game the reward model into emitting high rewards by visiting incorrectly classified states.

\noindent\textbf{Analysis.}
The results in Figure \ref{fig:rewhack} show that our method drives down the false positive rate in 2D navigation: the learned reward model rarely incorrectly classifies an unsafe or neutral state as a good state.
As a result, the deployed agent actually performs the desired task (center plot in Figure \ref{fig:robustness}), instead of seeking false positives.
As discussed in Section \ref{exp-safeexp} and illustrated in the right-most plot of Figure \ref{fig:safeexp}, the baselines learn a reward model that incorrectly extrapolates that continuing up and to the right past the goal region is good behavior.

For a concrete example of reward-maximizing synthetic queries that detect reward hacking, consider the reward-maximizing queries in the upper right corner of Figure \ref{fig:pointmass}, which are analyzed in Section \ref{ablation}.

\subsection{Safe Exploration} \label{exp-safeexp}

\begin{figure*}[t]
    \centering
    \includegraphics[height=0.16\textheight]{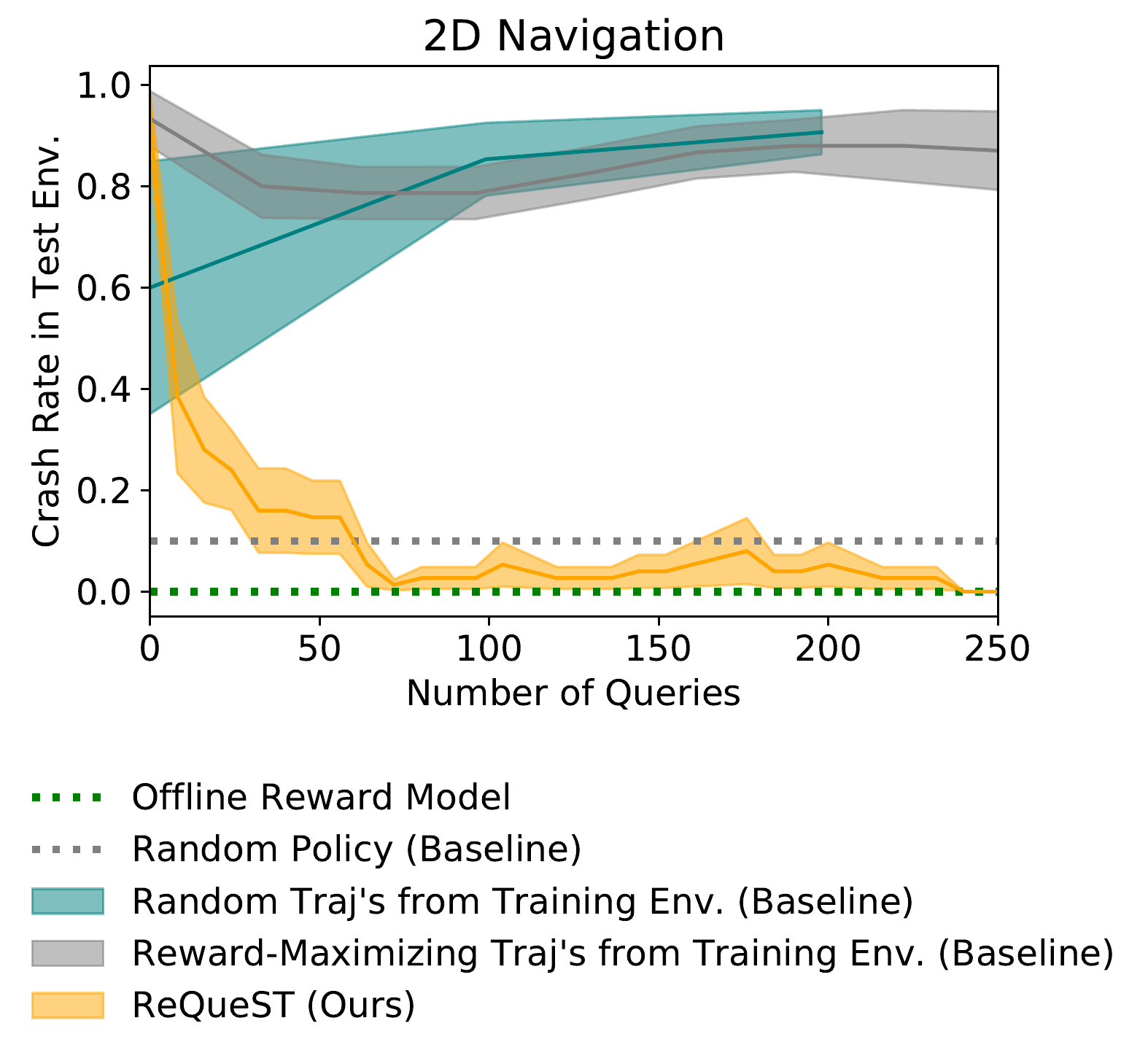}
    \includegraphics[height=0.16\textheight]{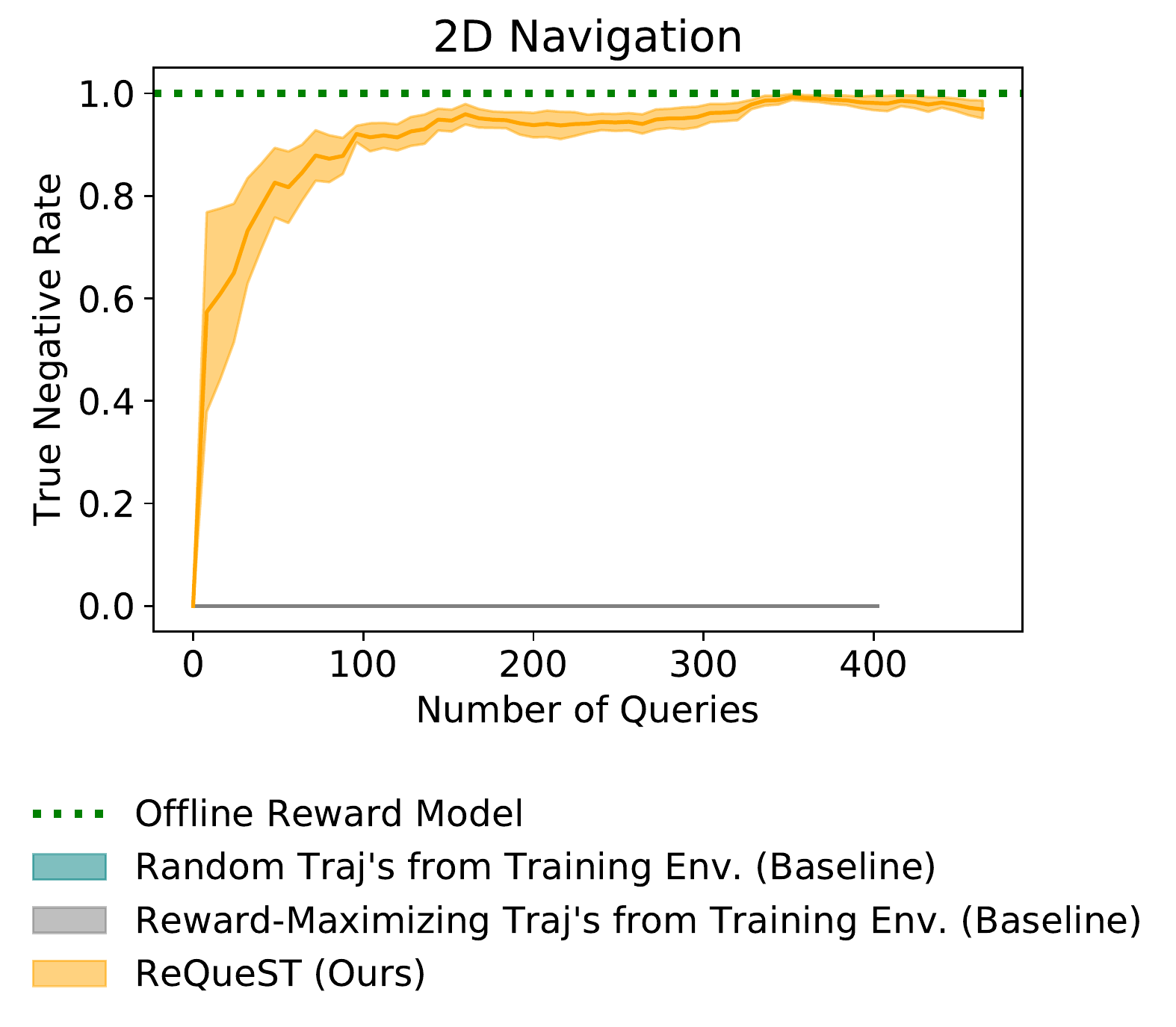}
    \includegraphics[height=0.16\textheight]{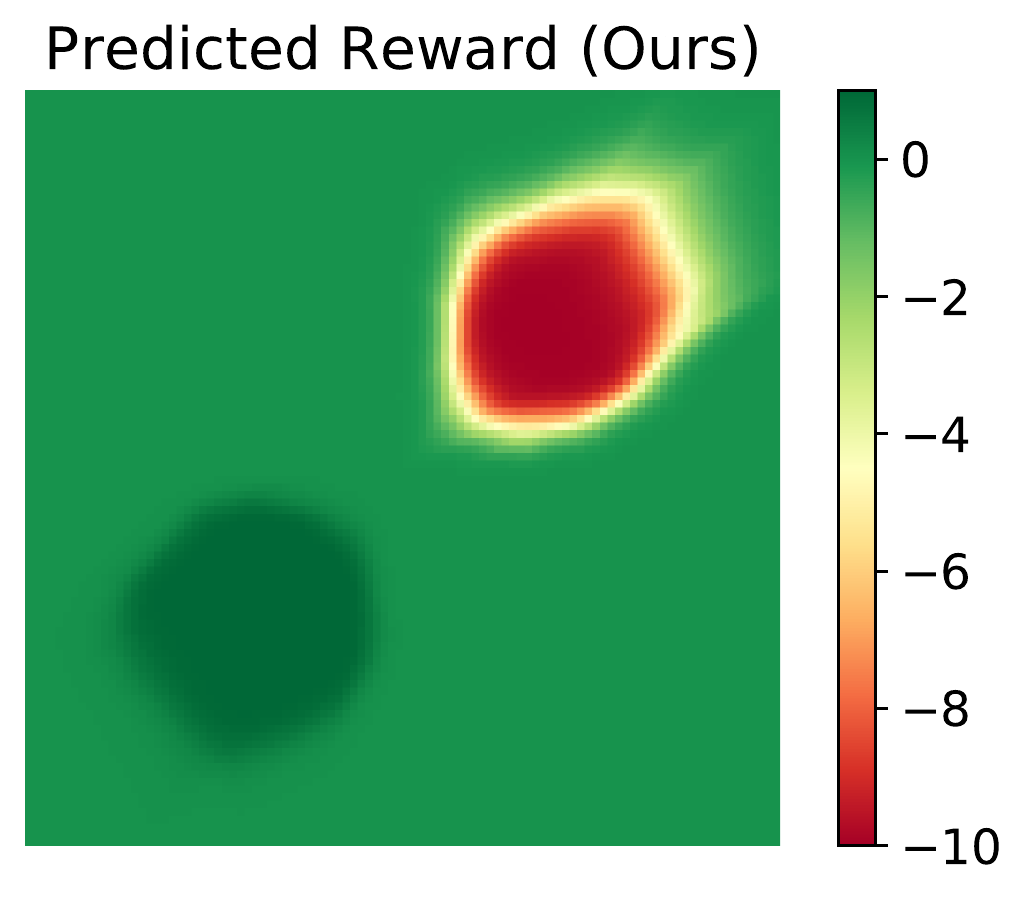}
    \includegraphics[height=0.16\textheight]{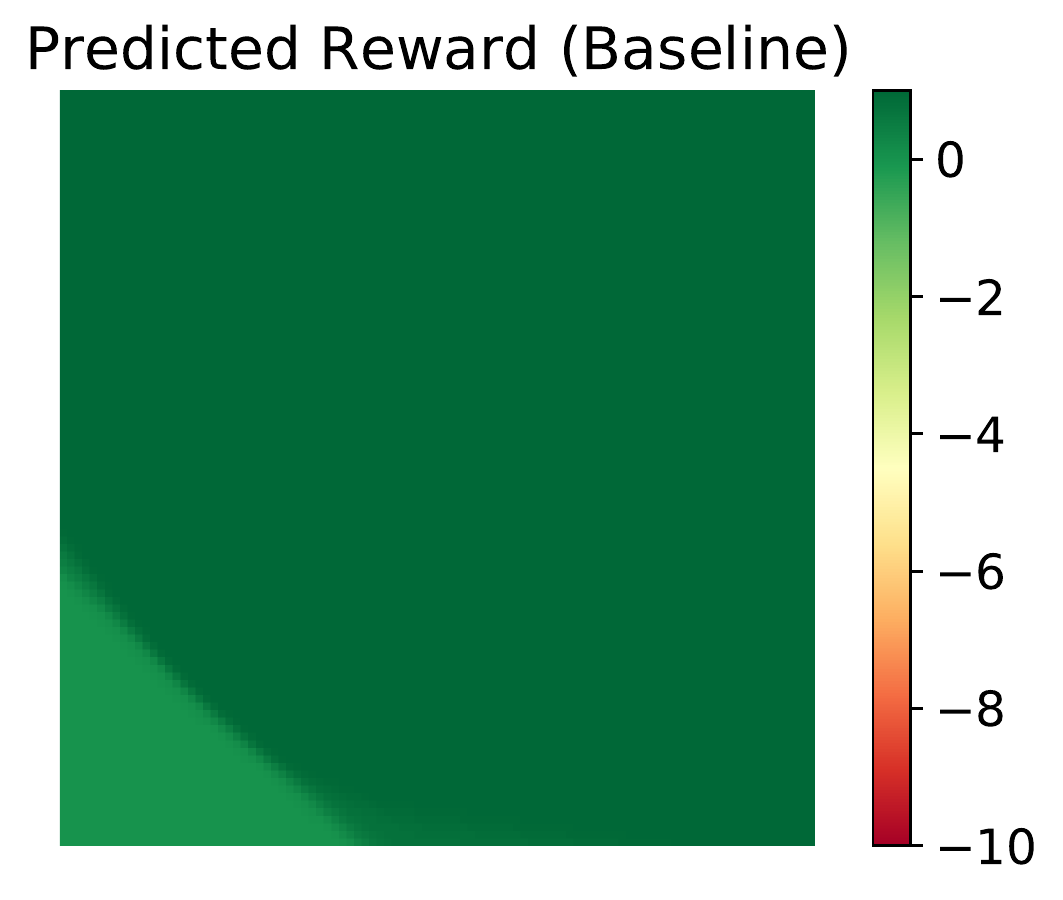}
    \caption{Experiments that address \textbf{Q3} -- can our method safely learn about unsafe states? -- by comparing our method, which uses synthetic trajectories, to baselines that only use real trajectories generated in the training environment. The results on 2D navigation show that our method (orange) significantly outperforms the baselines (blue and gray). The x-axis represents the number of queries to the user, where each query elicits a label for a single state transition $(s, a, s')$. The shaded areas show standard error over three random seeds. The heat maps represent the reward models learned by our method (left) and by the baselines (right).}
    \label{fig:safeexp}
\end{figure*}

\begin{figure*}[t]
    \centering
    \includegraphics[height=0.22\textheight]{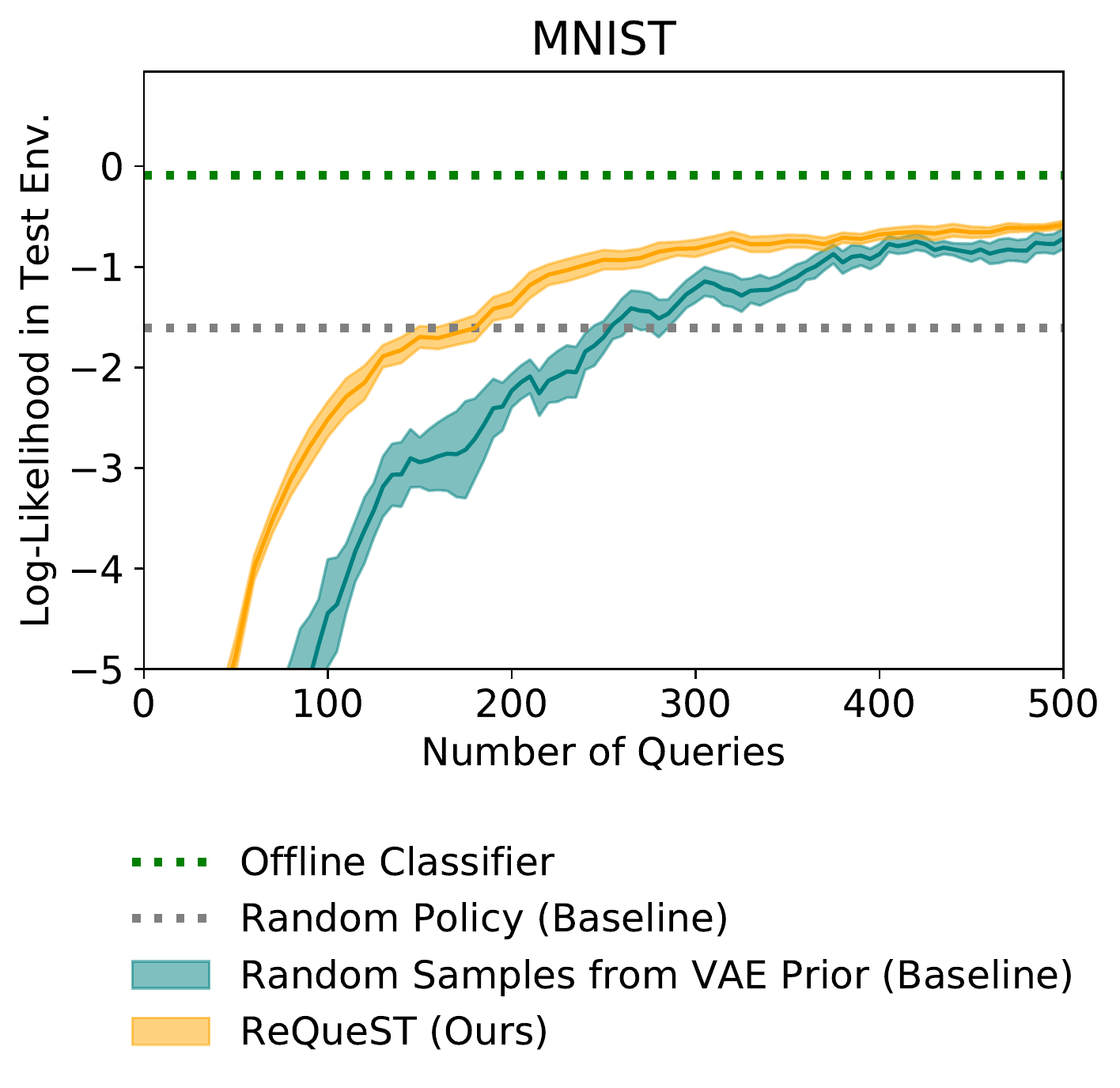}
    \includegraphics[height=0.22\textheight]{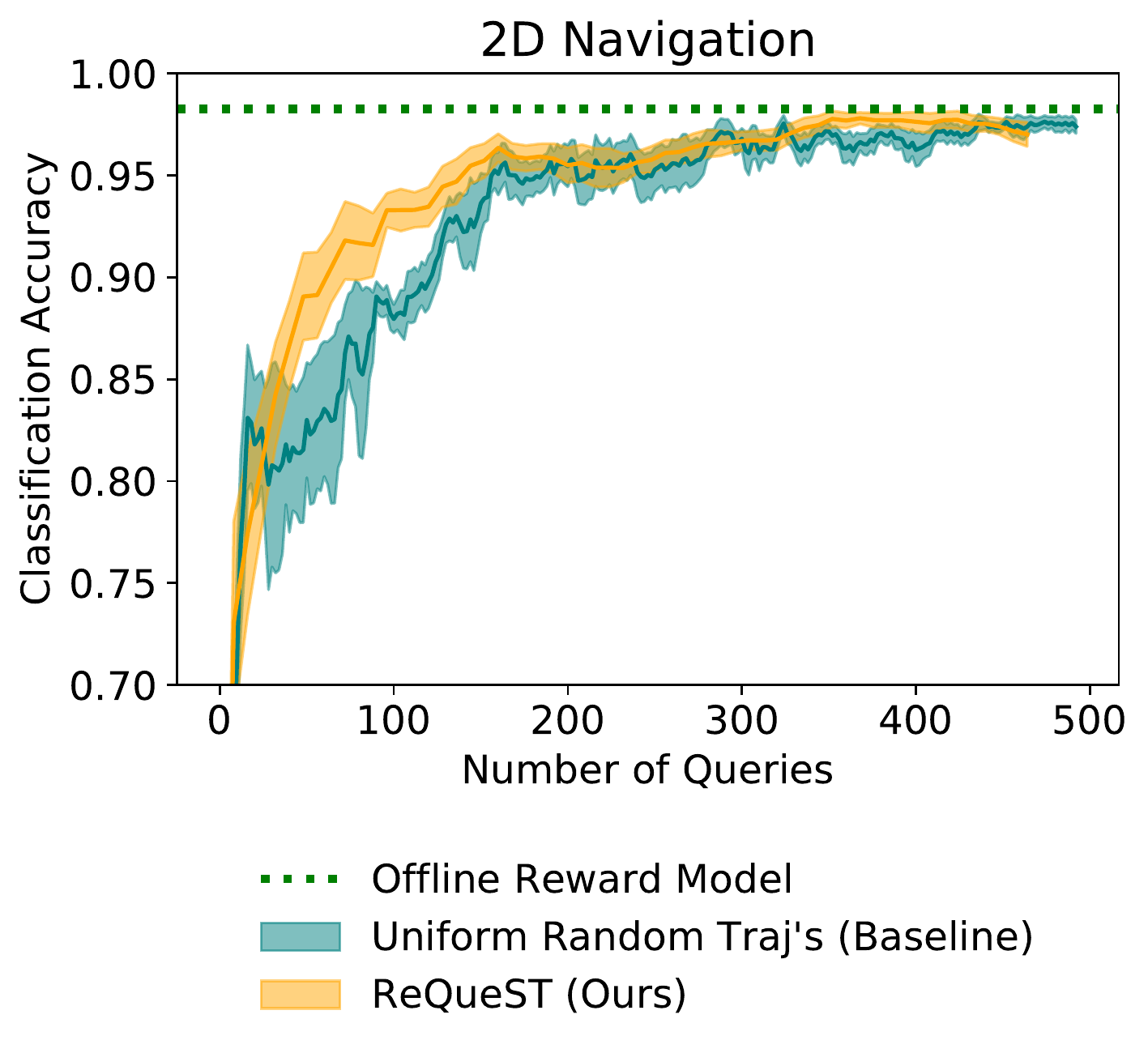}
    \includegraphics[height=0.22\textheight]{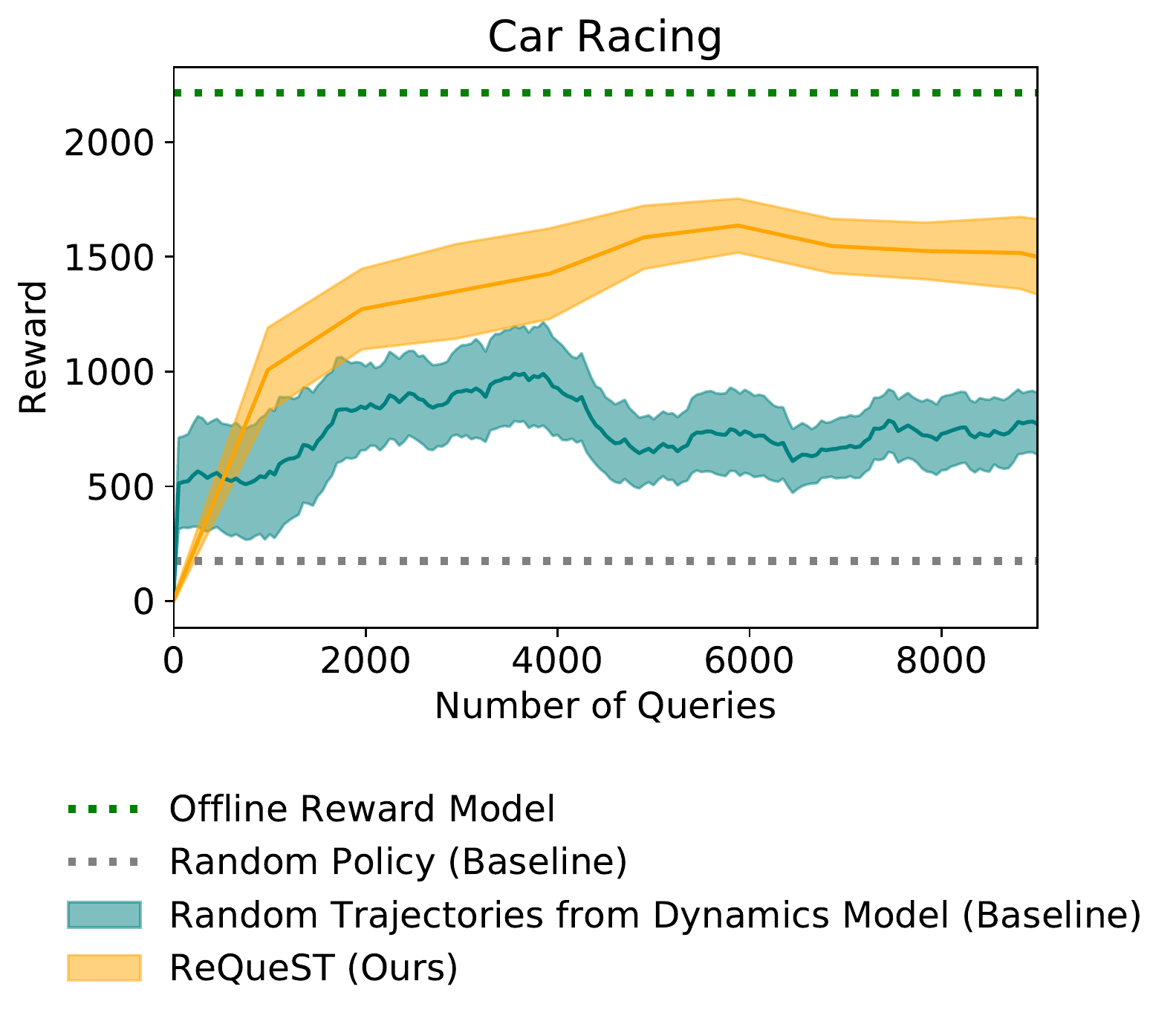}
    \caption{Experiments that address \textbf{Q4} -- do the proposed AFs improve upon random sampling from the generative model? -- by comparing our method, which synthesizes trajectories by optimizing AFs, to a baseline that ignores the AFs and randomly samples from the generative model. The results on MNIST, 2D navigation, and Car Racing show that our method (orange) significantly outperforms the baseline (blue) in Car Racing, and learns faster in MNIST and 2D navigation. The x-axis represents the number of queries to the user, where each query elicits a label for a single state transition $(s, a, s')$. The shaded areas show standard error over three random seeds.}
    \label{fig:query-efficiency}
\end{figure*}

One of the benefits of our method is that it can learn a reward model that accurately detects unsafe states, without having to visit unsafe states during the training process.
In the next experiment, we test this claim.

\noindent\textbf{Manipulated factors.}
We replicate the experimental setup in Section \ref{exp-robustness} for 2D navigation, including the same baselines.

\noindent\textbf{Dependent measures.}
We measure performance using the true negative rate of the reward model: the fraction of unsafe states correctly classified as unsafe, evaluated on the offline dataset of trajectories described in Section \ref{exp-robustness}.
We also use the crash rate of the deployed agent: the rate at which it gets caught in the trap region.

\noindent\textbf{Analysis.}
The results in Figure \ref{fig:safeexp} show that our method learns a reward model that classifies all unsafe states as unsafe, without visiting unsafe states during training (second and third figure from left); in fact, without visiting any states at all, since the queries are synthetic.
This enables the agent to avoid crashing during deployment (first figure from left).
The baselines differ from our method in that they actually have to visit unsafe states in order to query the user for labels at those states.
Since the baselines tend to not visit unsafe states during training, they do not learn about unsafe states (second and fourth figure from left), and the agent frequently crashes during deployment (first figure from left).

\subsection{Query Efficiency Compared to Baselines} \label{exp-queryeff}

\begin{figure*}[t]
    \centering
    \includegraphics[height=0.22\textheight]{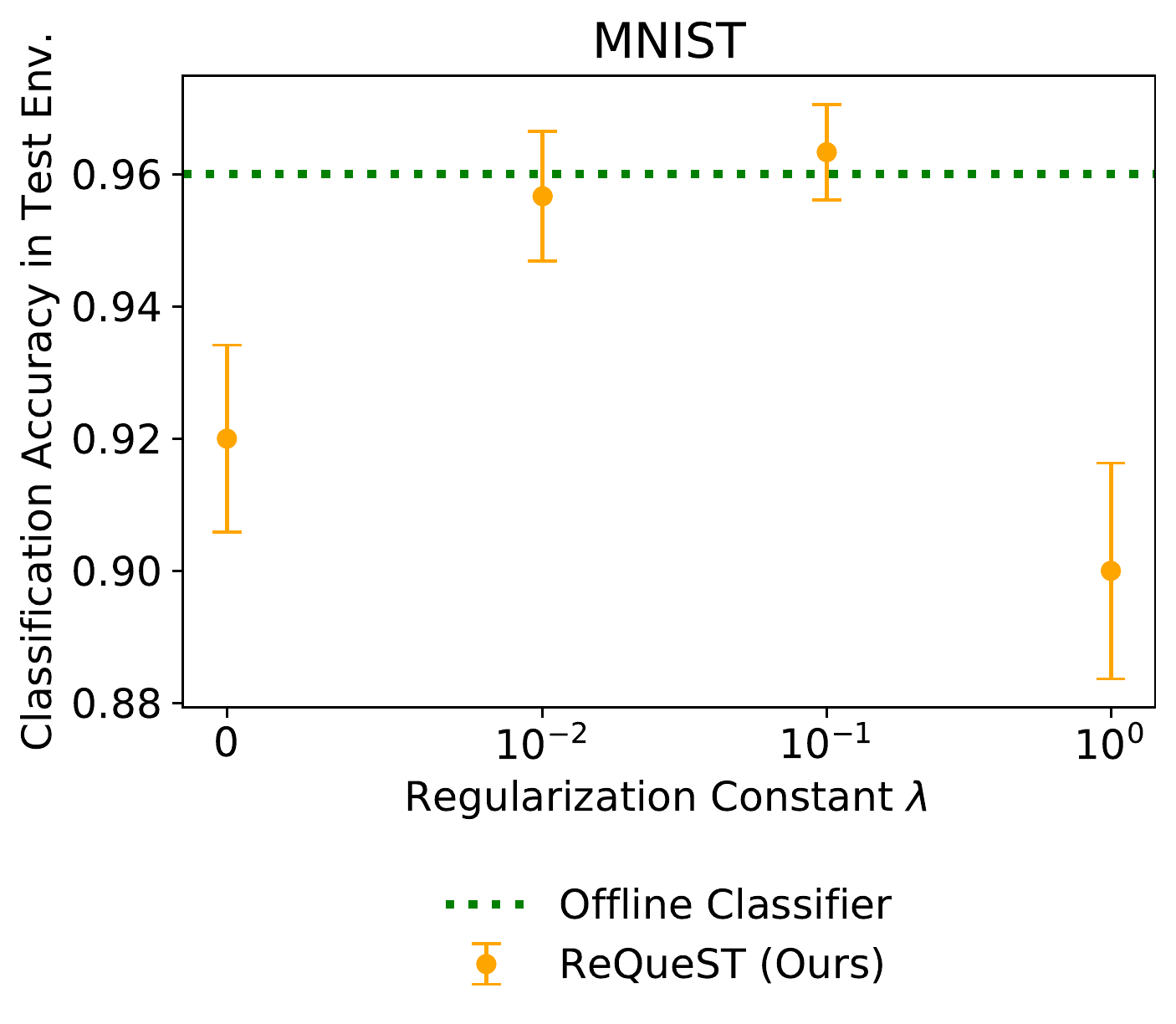}
    \includegraphics[height=0.22\textheight]{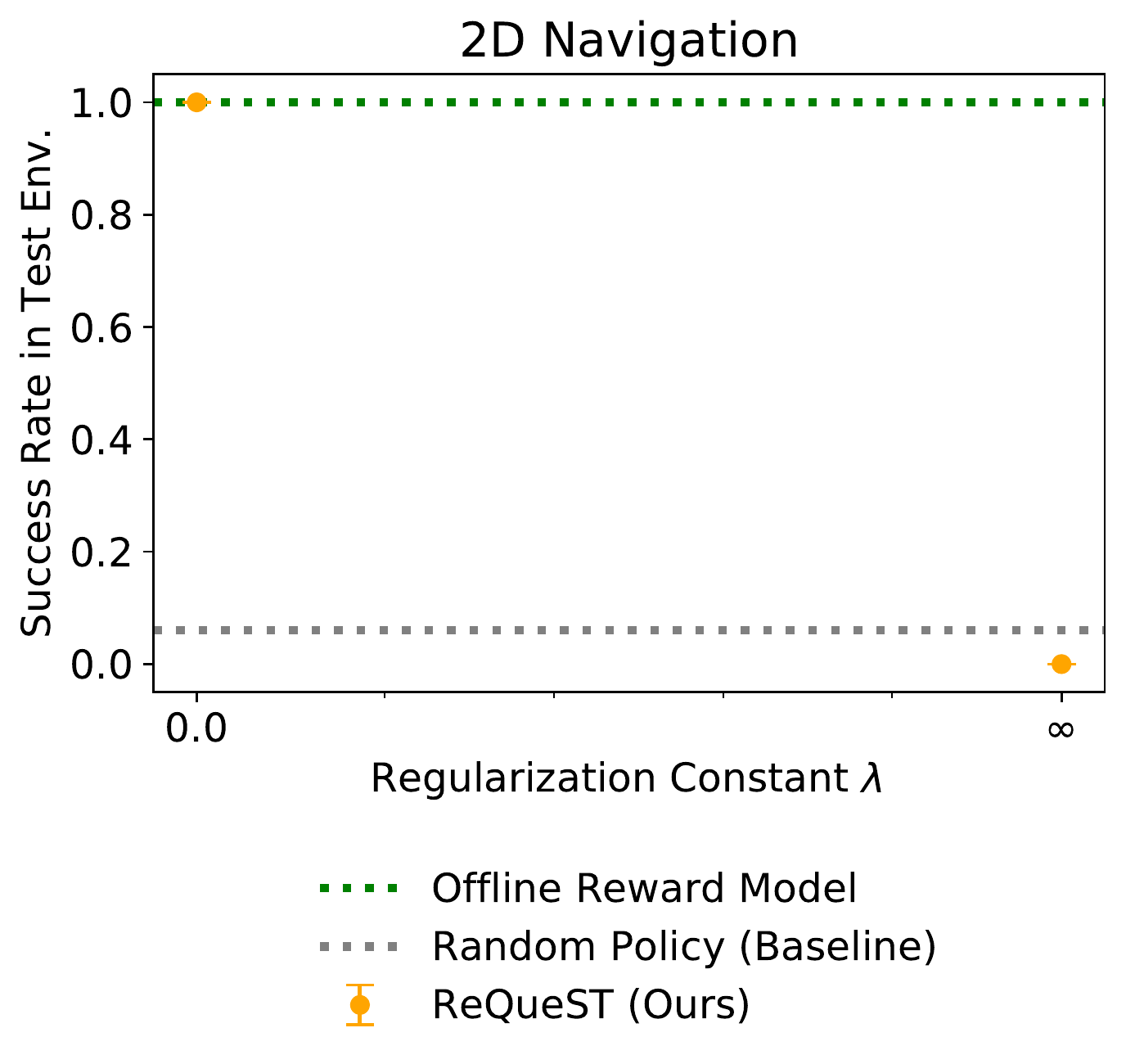}
    \includegraphics[height=0.22\textheight]{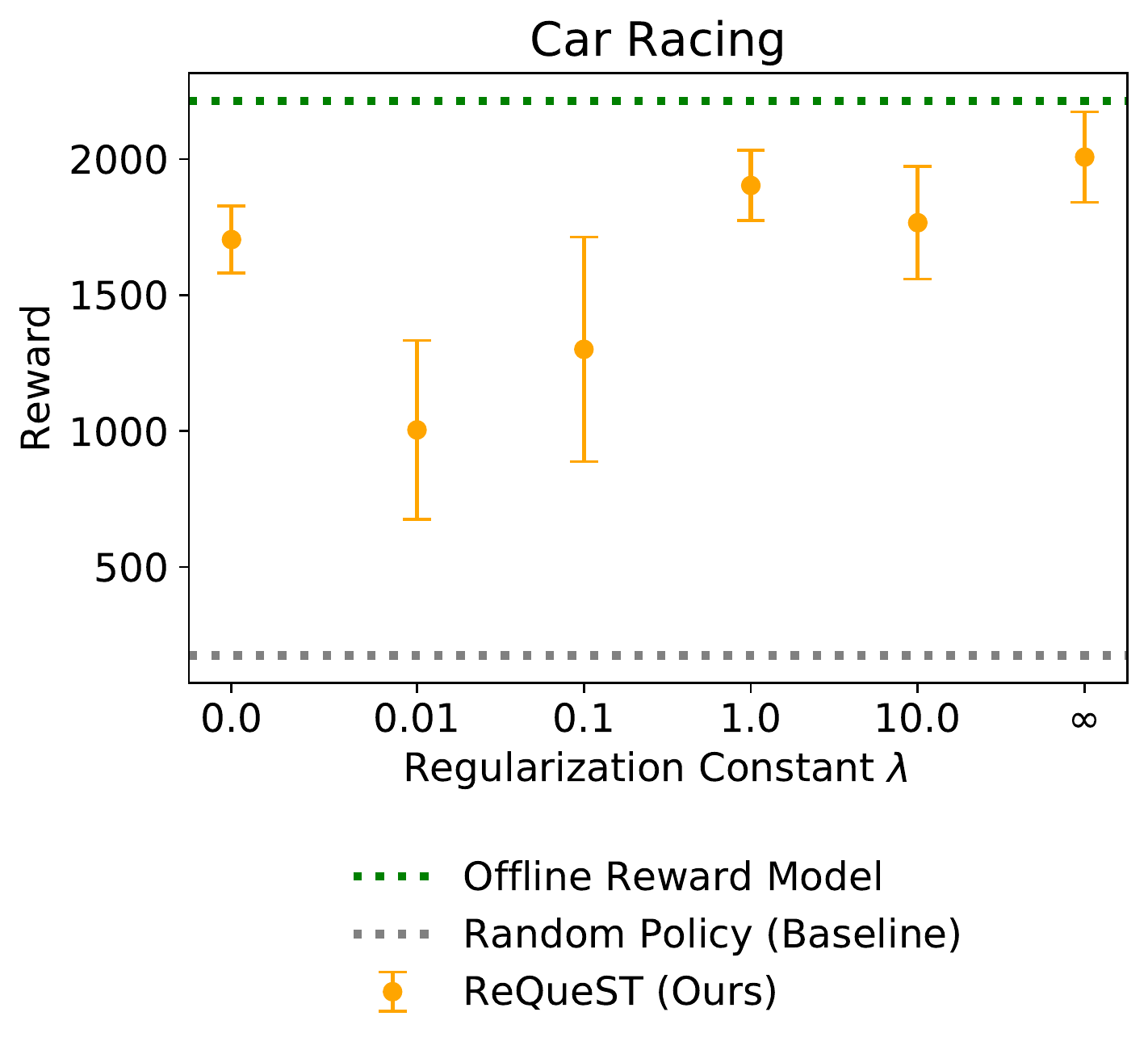}
    \caption{Experiments that address \textbf{Q5} -- how does the regularization constant $\lambda$ control the trade-off between realistic and informative queries? -- by evaluating our method with different values of $\lambda$, which controls the trade-off between producing realistic trajectories (higher $\lambda$) and informative trajectories (lower $\lambda$). The results on MNIST, 2D navigation, and Car Racing show that, while intermediate and low values of $\lambda$ work best for MNIST and 2D navigation respectively, a high value of $\lambda = \infty$ works best for Car Racing. The x-axis is log-scaled. The error bars show standard error over three random seeds, which is negligible in the results for 2D navigation.}
    \label{fig:lambda-sweep}
\end{figure*}

The previous experiment compared to baselines that are restricted to generating query trajectories by taking actions in the training environment.
In this experiment, we lift this restriction on the baselines: instead of taking actions in the training environment, the baselines can make use of the generative model trained in step (1).

\noindent\textbf{Manipulated factors.}
To answer \textbf{Q4}, we compare our method to a baseline that randomly samples trajectories from the generative model $p_{\bm{\phi}}$ -- using uniform random actions in Car Racing, samples from the VAE prior in MNIST, and uniform positions across the map in 2D navigation.

\noindent\textbf{Dependent measures.}
We measure performance in MNIST using the reward model's predicted log-likelihood of the ground-truth user labels in the test environment; in 2D navigation, the reward model's classification accuracy on an offline dataset containing states sampled uniformly throughout the environment; and in Car Racing, the true reward collected by an MPC agent that optimizes the learned reward, where the true reward gives a bonus for driving onto new patches of road, and penalizes going off-road.

\noindent\textbf{Analysis.}
The results in Figure \ref{fig:query-efficiency} show that our method, which optimizes trajectories using various AFs, requires fewer queries to the user than the baseline, which randomly samples trajectories.
This suggests that our four AFs guide query synthesis toward informative trajectories.
These results, and the results from Section \ref{exp-robustness}, suggest that our method benefits not only from using a generative model instead of the default training environment, but also from optimizing the AFs instead of randomly sampling from the generative model.

\subsection{Effect of Regularization Constant $\lambda$} \label{sweep}

One of the core features of our method is that, in Equation \ref{eq:query-synth}, it can trade off between producing realistic queries that maximize the regularization term $\log{p_{\bm{\phi}}(\tau)}$, and producing informative queries that maximize the AF $J(\tau)$.
In this experiment, we examine how the regularization constant $\lambda$ controls this trade-off, and how the trade-off affects performance.\footnote{Note that the scale of the optimal $\lambda$ may depend on the scale of the AF. In our experiments, we find that the same value of $\lambda$ generally works well for all four AFs. See Section \ref{imp-deets} in the appendix for details.}

\noindent\textbf{Manipulated factors.}
To answer \textbf{Q5}, we sweep different values of the regularization constant $\lambda$.
At one extreme, we constrain the query trajectories $\tau_{\text{query}}$ to be feasible under the generative model, by setting the next states $z_{t+1}$ to be the next states predicted by the dynamics model instead of free variables -- we label this setting as $\lambda = \infty$ for convenience (see Section \ref{imp-deets} in the appendix for details).
At the other extreme, we set $\lambda = 0$, which allows $\tau_{\text{query}}$ to be infeasible under the model.
Note that, even when $\lambda = 0$, the optimized trajectory $\tau_{\text{query}}$ is still regularized by the fact that it is optimized in the latent space of the state encoder $f$, instead of, e.g., raw pixel space.

\noindent\textbf{Dependent measures.}
We measure performance as in Section \ref{exp-robustness}.

\begin{figure*}[t]
    \centering
    \includegraphics[height=0.22\textheight]{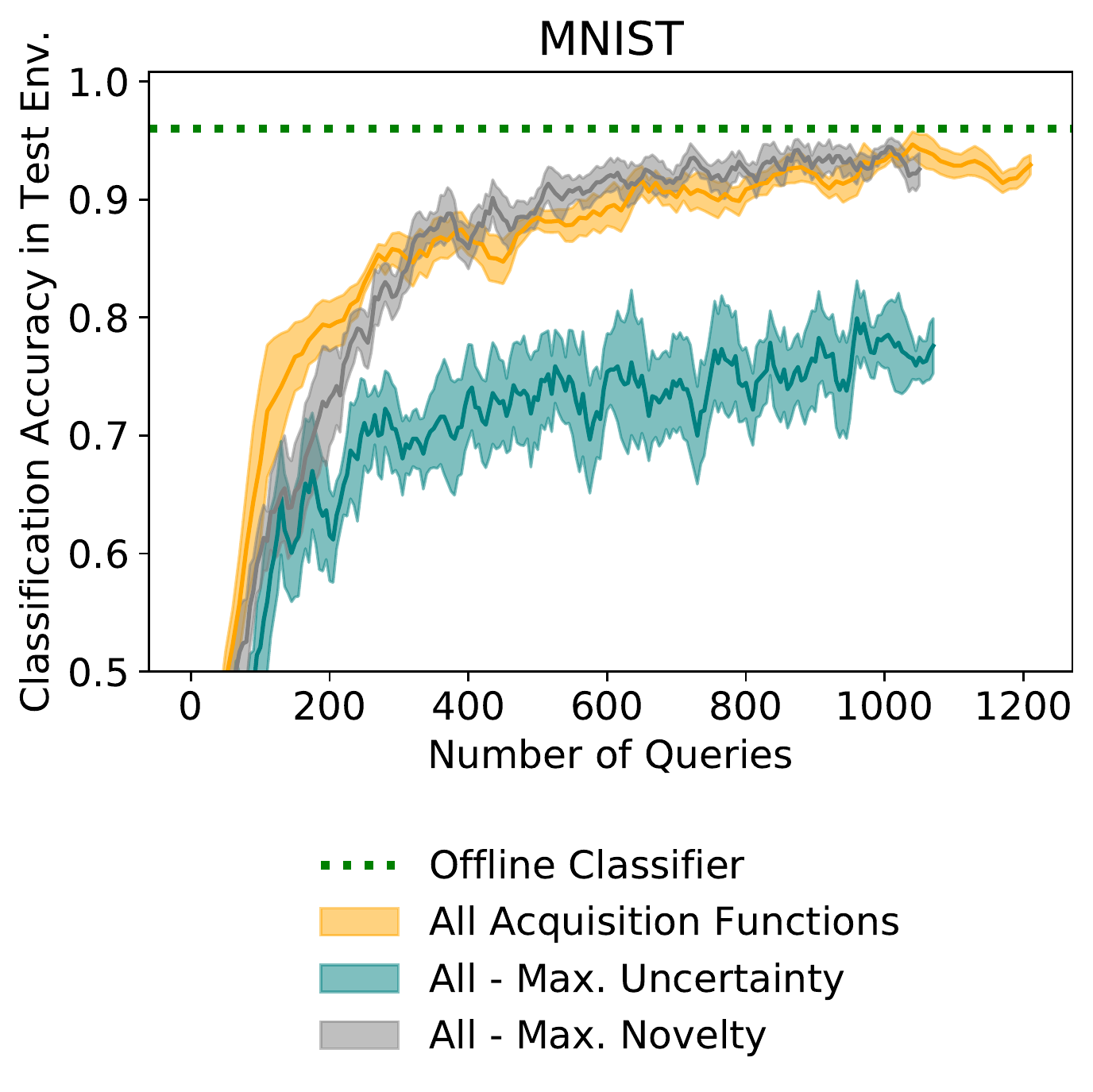}
    \includegraphics[height=0.22\textheight]{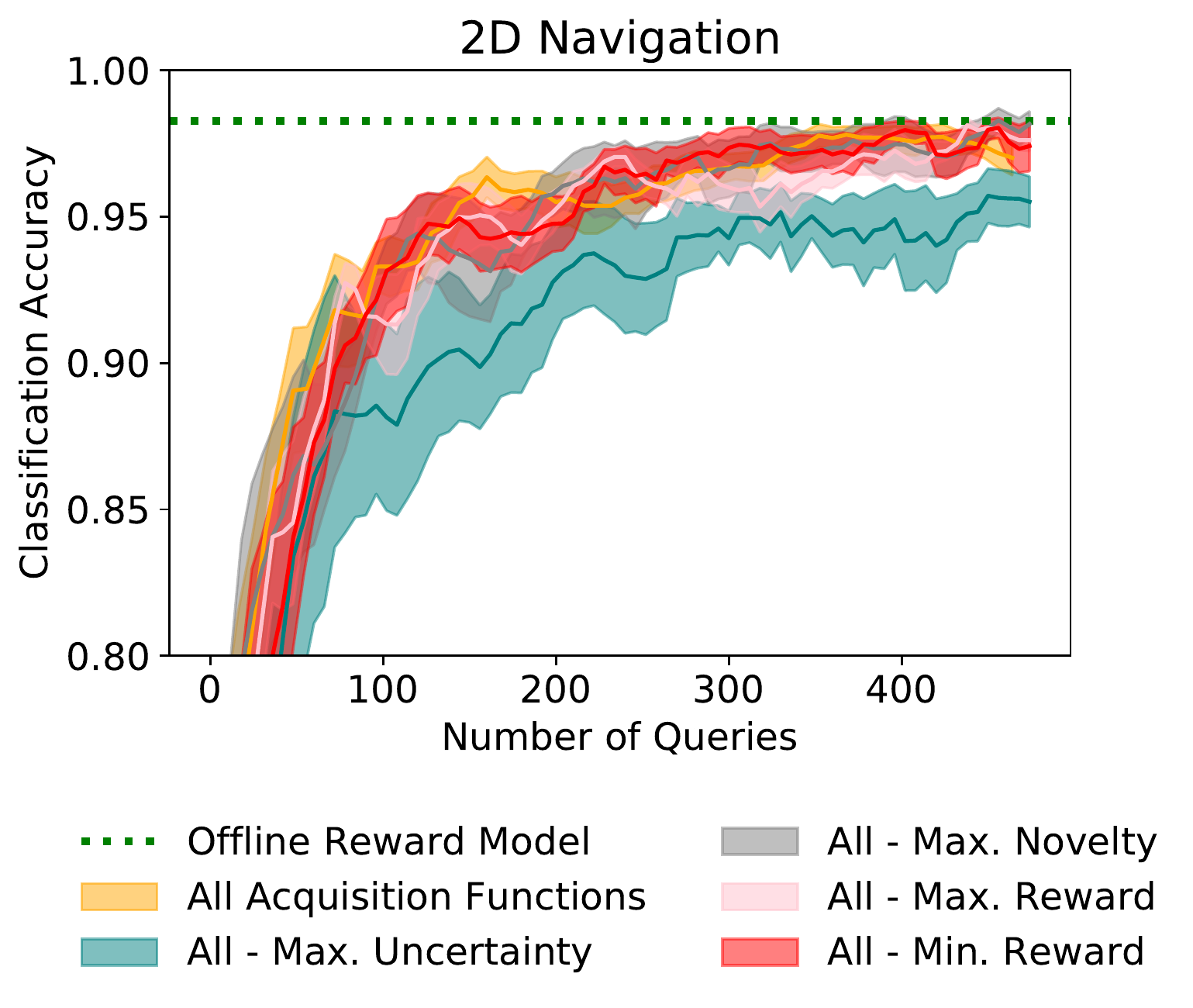}
    \includegraphics[height=0.22\textheight]{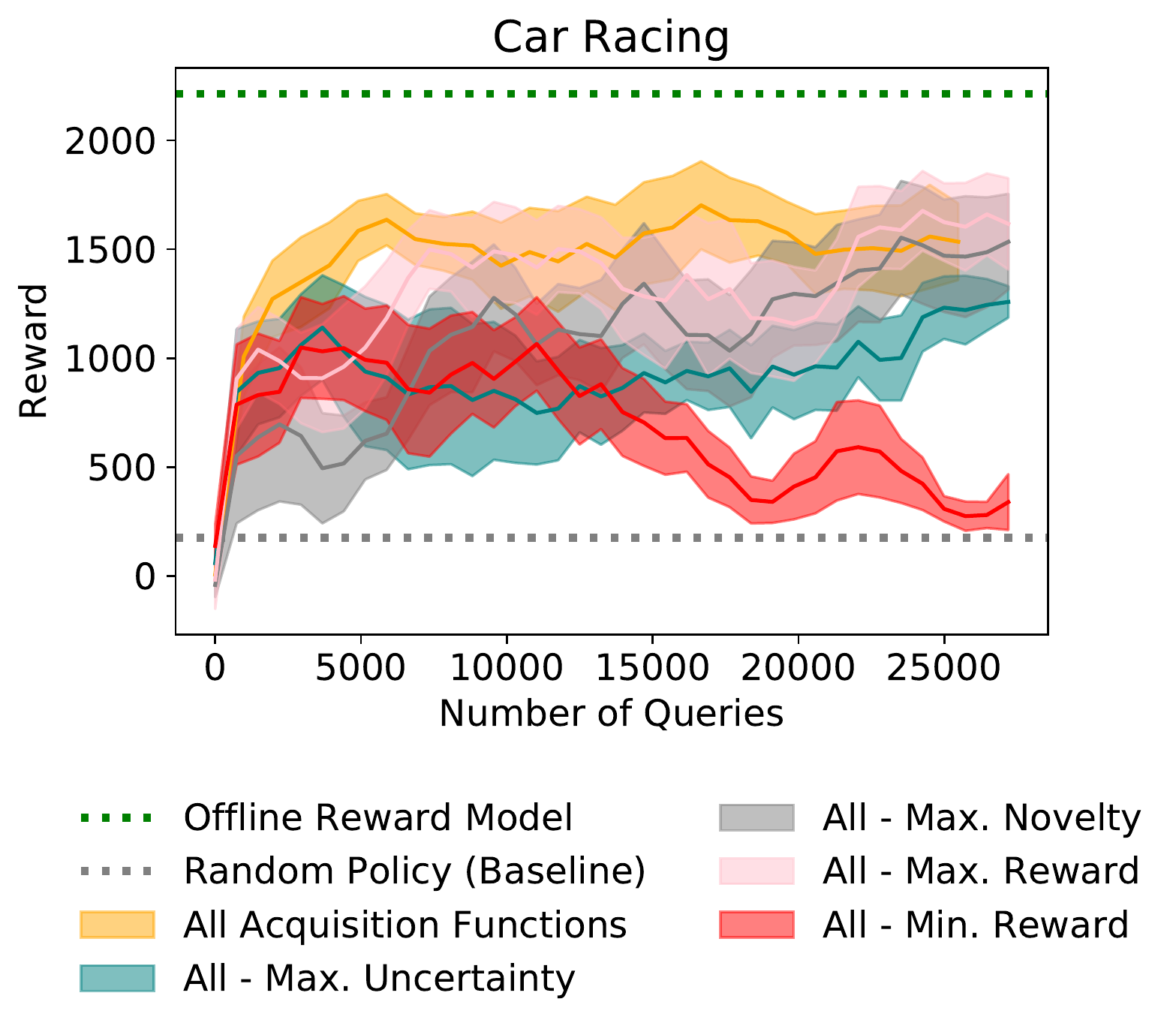}
    \caption{Experiments that address \textbf{Q6} -- how much do each of the four AFs contribute to performance? -- by comparing our method to ablated variants that drop each AF, one at a time, from the set of four AFs in line 4 of Algorithm \ref{alg:rqst-alg}. The results on MNIST, 2D navigation, and Car Racing show that our method (orange) generally outperforms its ablated variants (blue, gray, red, and pink), although the usefulness of each AF depends on the domain and amount of training data.. The x-axis represents the number of queries to the user, where each query elicits a label for a single state transition $(s, a, s')$. The shaded areas show standard error over three random seeds.}
    \label{fig:ablation}
\end{figure*}

\noindent\textbf{Analysis.}
The results in Figure \ref{fig:lambda-sweep} show that the usefulness of generating unrealistic trajectories depends on the domain.
In MNIST, producing unrealistic images by decreasing $\lambda$ can improve performance, although an intermediate value works best.
In 2D navigation, setting $\lambda$ to a low value is critical for learning the task. Note that we only tested $\lambda = 0$ and $\lambda = \infty$ in this domain, since we intentionally setup the training and test environments as a sanity check, where $\lambda = 0$ should perform best, and $\lambda = \infty$ should not succeed.
In Car Racing, constraining the queries to be feasible ($\lambda = \infty$) performs best.

There is a trade-off between being informative (by maximizing the AF) and staying on the distribution of states in the training environment (by maximizing likelihood).
In domains like Car Racing -- where the training and test environments have similar state distributions, and off-distribution queries can be difficult for the user to interpret and label -- it makes sense to trade off being informative for staying on-distribution.
In domains like MNIST and 2D navigation, where we intentionally create a significant shift in the state distribution between the training and test environments, it makes more sense to trade off staying on-distribution for being informative.

\noindent\textbf{Visualizing synthesized queries.}
Figure \ref{fig:carracing} in the appendix shows examples of Car Racing query trajectories $\tau_{\text{query}}$ optimized with either $\lambda = 0$ or $\lambda = \infty$.
Unsurprisingly, the $\lambda = 0$ queries appear less realistic, but clearly maximize the AF better than their $\lambda = \infty$ counterparts.

\subsection{Acquisition Function Ablation Study} \label{ablation}

We propose four AFs intended to produce different types of hypotheticals.
In this experiment, we investigate the contribution of each type of query to the performance of the overall method.

\noindent\textbf{Manipulated factors.}
To answer \textbf{Q6}, we conduct an ablation study, in which we drop out each the four AFs, one by one, from line 4 in Algorithm \ref{alg:rqst-alg}, and measure the performance of only generating queries using the remaining three AFs.
We also visualize the queries generated by each of the four AFs, to illustrate their qualitative properties.

\noindent\textbf{Dependent measures.}
We measure performance as in Section \ref{exp-queryeff}.

\noindent\textbf{Analysis.}
The results in Figure \ref{fig:ablation} show that the usefulness of each AF depends on the domain and the amount of training data collected.

In MNIST, dropping $J_u$ hurts performance, suggesting that uncertainty-maximizing queries elicit useful labels.
Dropping $J_n$ also hurts performance when the number of queries is small, but actually improves performance if enough queries have already been collected.
Novelty-maximizing queries tend to be repetitive in practice: although they are distant from the training data in terms of Equation \ref{eq:dist}, they are visually similar to the existing training data in that they appear to be the same digits.
Hence, while they are helpful at first, they hurt query efficiency later in training.

In 2D navigation, dropping $J_u$ hurts performance, while dropping any of the other AFs individually does not hurt performance.
These results suggest that uncertainty-maximizing queries can be useful, in domains like MNIST and 2D navigation, where uncertainty can be modeled and estimated accurately.

In Car Racing, dropping $J_-$ hurts the most.
Reward-minimizing queries elicit labels for unsafe states, which are rare in the training environment unless you explicitly seek them out.
Hence, this type of query performs the desired function of augmenting the training data with more examples of unsafe states, thereby making the reward model better at detecting unsafe states.

\noindent\textbf{Visualizing synthesized queries.}
Figure \ref{fig:pointmass} and Figures \ref{fig:mnist} and \ref{fig:carracing} in the appendix illustrate examples of queries generated by each of the four AFs.

In MNIST (Figure \ref{fig:mnist} in the appendix), the uncertainty-maximizing queries are digits that appear ambiguous but coherent, while the novelty-maximizing queries tend to cluster around a small subset of the digits and appear grainy.

In 2D navigation (Figure \ref{fig:pointmass}), the demonstrations contain mostly neutral states en route to the goal, and a few good states at the goal.
If we were to train on only the demonstrations, the reward model would be unaware of the trap.
Initially, the queries, which we restrict to just one state transition from the initial state $s_0$ to a synthesized next state $s_1$, are relatively uniform.
The first reward-maximizing queries are in the upper right corner, which makes sense: the demonstrations contain neutral states in the lower left, and good states farther up and to the right inside the goal region, so the reward model extrapolates that continuing up and to the right, past the goal region, is good behavior.
The reward model, at this stage, is susceptible to reward hacking -- a problem that gets addressed when the user labels the reward-maximizing queries in the upper right corner as neutral.

After a few more queries, the reward-maximizing queries start to cluster inside the goal region, and the reward-minimizing queries cluster inside the trap.
This is helpful early during training, for identifying the locations of the goal and trap.
The uncertainty-maximizing queries cluster around the boundaries of the goal and the trap, since that is where model uncertainty is highest.
This is helpful for refining the reward model's knowledge of the shapes of the goal and trap.
The novelty-maximizing queries get pushed to the corners of the environment.
This is helpful for determining that the goal and trap are relatively small and circular, and do not bleed into the corners of map.

In Car Racing (Figure \ref{fig:carracing} in the appendix), the reward-maximizing queries show the car driving down the road and making a turn.
The reward-minimizing queries show the car going off-road as quickly as possible.
The uncertainty-maximizing queries show the car driving to the edge of the road and slowing down.
The novelty-maximizing queries show the car staying still, which makes sense since the training data tends to contain mostly trajectories of the car in motion.

\section{Discussion}

\noindent\textbf{Summary.}
We contribute the ReQueST algorithm for learning a reward model from user feedback on hypothetical behavior.
The key idea is to automatically generate hypotheticals that efficiently determine the user's objectives.
Simulated experiments on MNIST, state-based 2D navigation, and image-based Car Racing show that our method produces accurate reward models that transfer well to new environments and require fewer queries to the user, compared to baseline methods adapted from prior work.
Our method detects reward hacking before the agent is deployed, and safely learns about unsafe states.
Through a hyperparameter sweep, we find that our method can trade off between producing realistic vs. informative queries, and that the optimal trade-off varies across domains.
Through an ablation study, we find that the usefulness of each of the four acquisition functions we propose for optimizing queries depends on the domain and the amount of training data collected.
Our experiments broadly illustrate how models of the environment can be used to improve the way we learn models of task rewards.

\noindent\textbf{Limitations and future work.}
The main practical limitation of our method is that it requires a generative model of initial states and a forward dynamics model, which can be difficult to learn from purely off-policy data in complex, visual environments.
One direction for future work is relaxing this assumption; e.g., by incrementally training a generative model on on-policy data collected from an RL agent in the training environment~\citep{kaiser2019model,hafner2018learning}.
Another direction is to address the safety concerns of training on unsupervised interactions by using safe expert demonstrations instead (as discussed in Section \ref{safety}).

Our method assumes the user can label agent behaviors with rewards.
For complex tasks that involve extremely long-term decision-making and high-dimensional state spaces, such as managing public transit systems or sequential drug treatments, the user may not be able to meaningfully evaluate the performance of the agent.
To address this issue, one could implement ReQueST inside a framework that enables users to evaluate complex behaviors, such as recursive reward modeling~\citep{leike2018scalable} or iterated amplification~\citep{christiano2018supervising}.

\section*{Acknowledgments}
Thanks to Gabriella Bensinyor, Tim Genewein, Ramana Kumar, Tom McGrath, Victoria Krakovna, Tom Everitt, Zac Kenton, Richard Ngo, Miljan Martic, Adam Gleave, and Eric Langlois for useful suggestions and feedback.
Thanks in particular to Gabriella Bensinyor, who proposed the name and acronym of our method: reward query synthesis via trajectory optimization, or ReQueST.
This work was supported in part by an NVIDIA Graduate Fellowship.

\bibliography{master}
\bibliographystyle{icml2019}

\clearpage

\appendix
\section{Appendix}

\subsection{Implementation Details} \label{imp-deets}

\noindent\textbf{Shooting vs. collocation.}
We use the notation $\lambda = \infty$ to denote solving the optimization problem in Equation \ref{eq:query-synth} with a shooting method instead of a collocation method. The shooting method optimizes $(z_0, a_0, a_1, ..., a_{T-1})$, and represents $z_{t+1} = \mathbb{E}_{\bm{\phi}}[z_{t+1} | z_0, a_0, a_1, ..., a_t]$ using the forward dynamics model $p_{\bm{\phi}}$ learned in step (1).

\noindent\textbf{MNIST classification.}
We simulate the user in line 7 of Algorithm \ref{alg:rqst-alg} as an expert, k-nearest neighbors classifier $p_{\text{user}}(a | s)$ trained on all labeled data.
We only generate queries using the AFs $J_n$ and $J_u$ in line 4 of Algorithm \ref{alg:rqst-alg}, since $J_+$ and $J_-$ are not useful for single-step classification.
We replace $p_{\bm{\theta}_i}(c | s, a, s')$ with $p_{\bm{\theta}_i}(a | s)$ in Equation \ref{eq:disag} and line 12 of Algorithm \ref{alg:rqst-alg}.
We represent $p_{\bm{\theta}}(a | s)$ in Equation \ref{eq:rew-clf} using a feedforward neural network with two fully-connected hidden layers containing 256 hidden units each, and $m = 4$ separate networks in the ensemble.
The MPC agent in Equation \ref{eq:mpc} reduces to $\pi_{\text{mpc}}(a | s) = p_{\bm{\theta}}(a | s)$.
The Gaussian prior distribution of the VAE yields the likelihood model, $p_{\bm{\phi}}(s_0) \propto \exp{(\|f_{\bm{\phi}}(s_0)\|_2^2)}$.
The state inputs to the reward model are the latent embeddings produced by $f_{\bm{\phi}}$, instead of the raw pixel inputs.
We set $\lambda = 0.1$ when synthesizing queries with the AF $J_u$, and $\lambda = 0.01$ when synthesizing queries with the AF $J_n$.

\noindent\textbf{State-based 2D navigation.}
To encourage the agent to avoid the trap, the reward constants are asymmetric: $R_{\text{good}} = 1$, $R_{\text{unsafe}} = -10$, and $R_{\text{neutral}} = 0$.
Since the states are already low-dimensional, we simply use the identity function for the state encoder and decoder.
We represent $p_{\bm{\theta}}(c | s, a, s')$ in Equation \ref{eq:rew-clf} using a feedforward neural network with two fully-connected hidden layers containing 32 hidden units each, and $m = 4$ separate networks in the ensemble.
We hard-code a Gaussian forward dynamics model, $p(s_{t+1} | s_t, a_t) = \mathcal{N}(s_{t+1}; s_t + a_t, \sigma^2)$.
Each episode lasts at most 1000 steps, and the maximum speed is restricted to $\|a\|_2 \leq 0.01$.
In Equation \ref{eq:mpc}, we use a planning horizon of $H = 500$.
In Equation \ref{eq:query-synth}, we use a query trajectory length of $T = 1$; i.e., the query consists of one state transition from the hard-coded initial state $s_0$ to a synthesized next state $s_1$.
We set $\lambda = 0$ when synthesizing queries for any of the four AFs.

\noindent\textbf{Car Racing.}
To encourage the agent to drive without being overly conservative, the reward constants are asymmetric: $R_{\text{good}} = 10$, $R_{\text{unsafe}} = -1$, and $R_{\text{neutral}} = 0$.
We represent $p_{\bm{\theta}}(c | s, a, s')$ in Equation \ref{eq:rew-clf} using a feedforward neural network with two fully-connected hidden layers containing 256 hidden units each, and $m = 4$ separate networks in the ensemble.
We train a generative model using the unsupervised approach in \citet{ha2018recurrent}, which learns a VAE state encoder and decoder with a 32-dimensional latent space, a recurrent dynamics model with a 256-dimensional latent space, and a mixture density network with 5 components that predicts stochastic transitions.
Since the environment is partially observable, we represent the state input to the reward model by concatenating the VAE latent embedding with the RNN latent embedding.
Each episode lasts at most 1000 timesteps.
In Equation \ref{eq:mpc}, we use a planning horizon of $H = 50$.
In Equation \ref{eq:query-synth}, we use a query trajectory length of $T = 50$.
We set $\lambda = \infty$ when synthesizing queries for any of the four AFs.

In the high-dimensional Car Racing environment, we find that optimizing Equation \ref{eq:query-synth} leads to incomprehensible query trajectories $\tau_{\text{query}}$, even for high values of the regularization constant $\lambda$.
To address this issue, we modify the method in two ways that provide additional regularization.
First, instead of optimizing the initial state $s_0$ in $\tau_{\text{query}}$, we set it to some real state sampled from the training environment during step (1).
Second, instead of optimizing $(z_0, a_0, z_1, ..., z_T)$, where $\tau = (f^{-1}(z_0), a_0, f^{-1}(z_1), a_1, ..., f^{-1}(z_T))$, we optimize $(z_0, a_0, m_0, a_1, m_1, ..., a_{T-1}, m_{T-1})$, where $\tau = (f^{-1}(z_0), a_0, f^{-1}(\mathrm{MDN}(z_0, a_0, m_0)), a_1$, ..., $ f^{-1}(\mathrm{MDN}(z_{T-1}, a_{T-1}, m_{T-1})))$. The function $\mathrm{MDN}(z_t, a_t, m_t)$ denotes using the mixture coefficients $m_t$ to compute the expected next state, instead of using the mixture coefficients $\psi(z_t, a_t)$ predicted by the mixture density network.
Thus, the likelihood regularization term becomes $\log{p_{\bm{\phi}}(\tau)} = \sum_{t=0}^{T-1} H(m_t, \psi(z_t, a_t))$, where $H$ is the cross-entropy.
This representation of the trajectory $\tau$ is easier to optimize, and results in more comprehensible queries.

\begin{figure}
    \centering
    \includegraphics[width=0.85\linewidth]{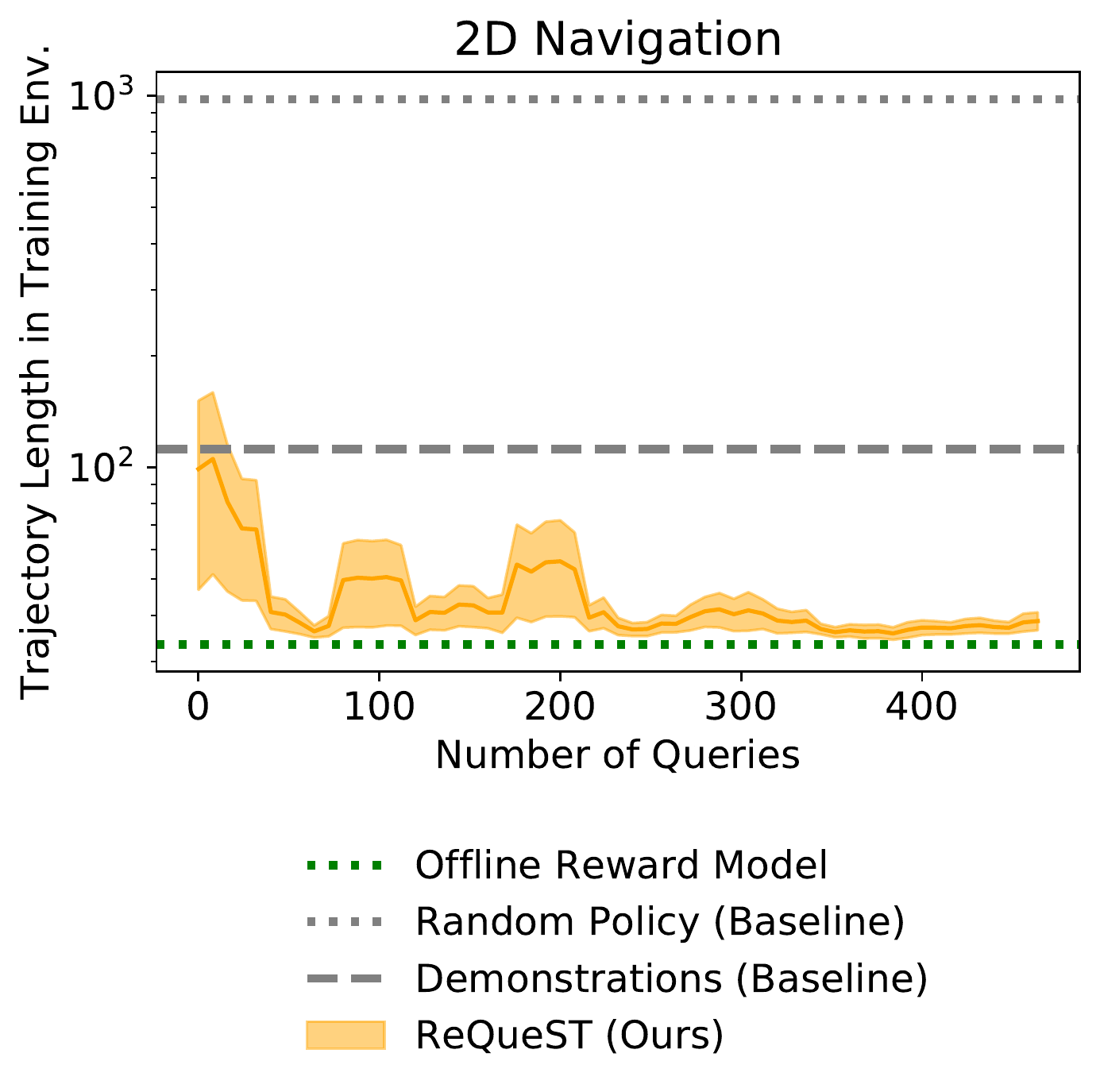}
    \includegraphics[width=0.85\linewidth]{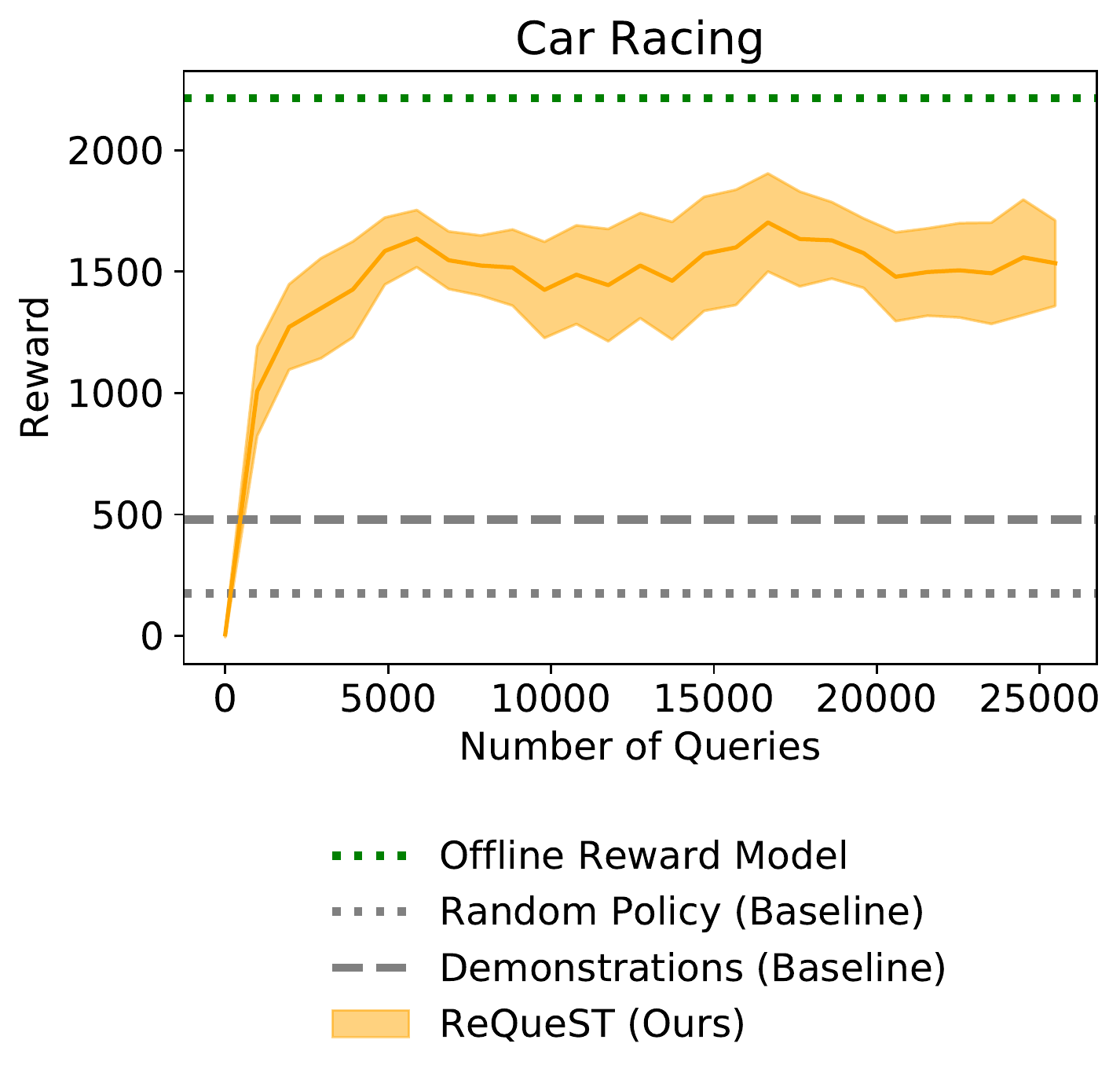}
    \caption{Our method initializes the reward model with suboptimal user demonstrations, in line 2 of Algorithm \ref{alg:rqst-alg}. The experiments in Section \ref{exp-robustness} show that our method learns a reward model that enables the agent to outperform the suboptimal demonstrator. In 2D navigation (top), the agent gets to the goal faster than the demonstrator, even in the training environment -- the demonstrator takes a tortuous path to the goal, while the agent goes straight to the goal. In Car Racing (bottom), the agent drives faster and visits more new road patches than the cautious, slow demonstrator. We do not include results for MNIST, since it does not make sense to initialize the classifier with incorrect labels in this domain.}
    \label{fig:superhuman}
\end{figure}

\begin{figure}
    \centering
    \includegraphics[width=0.85\linewidth]{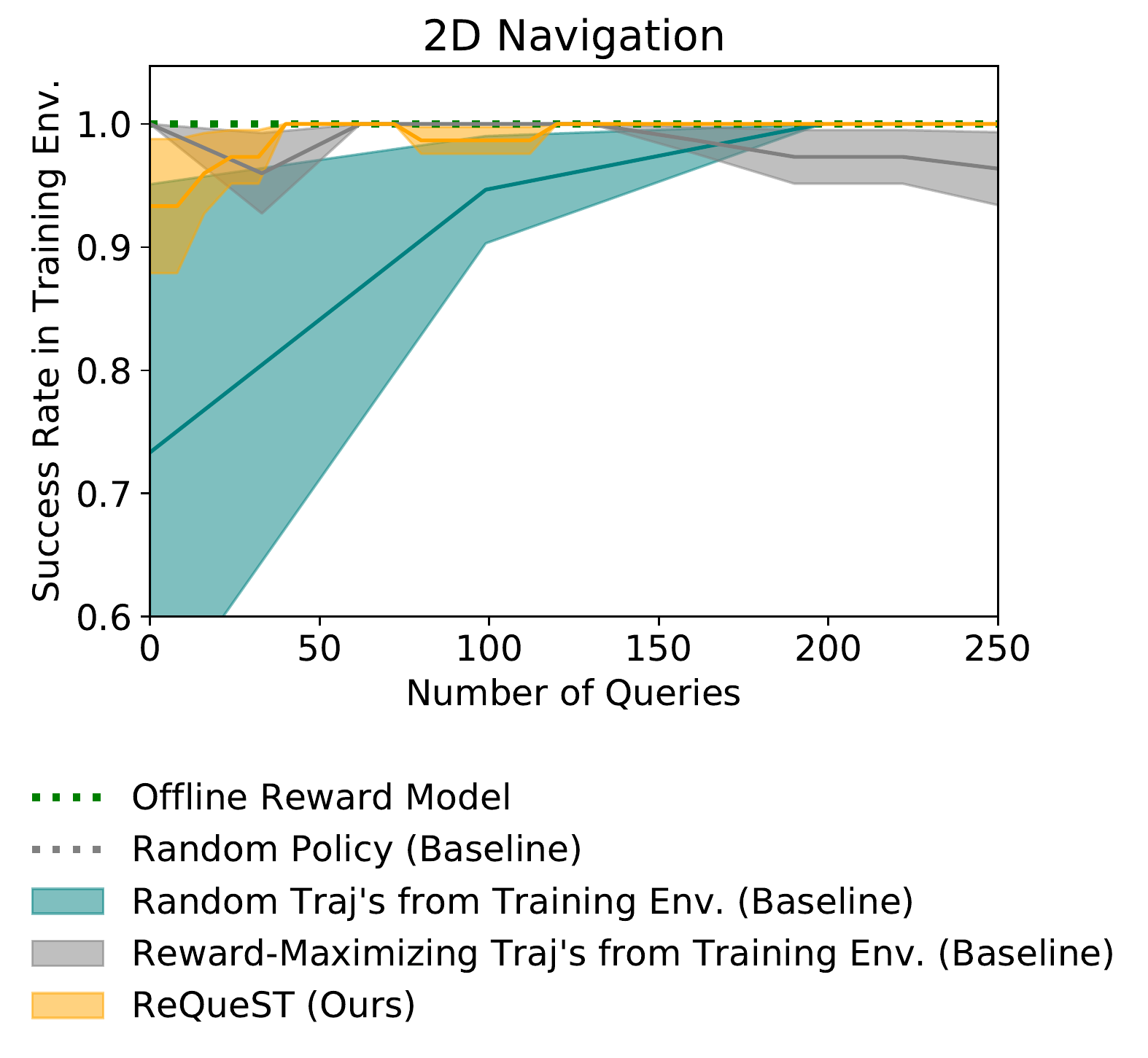}
    \includegraphics[width=0.85\linewidth]{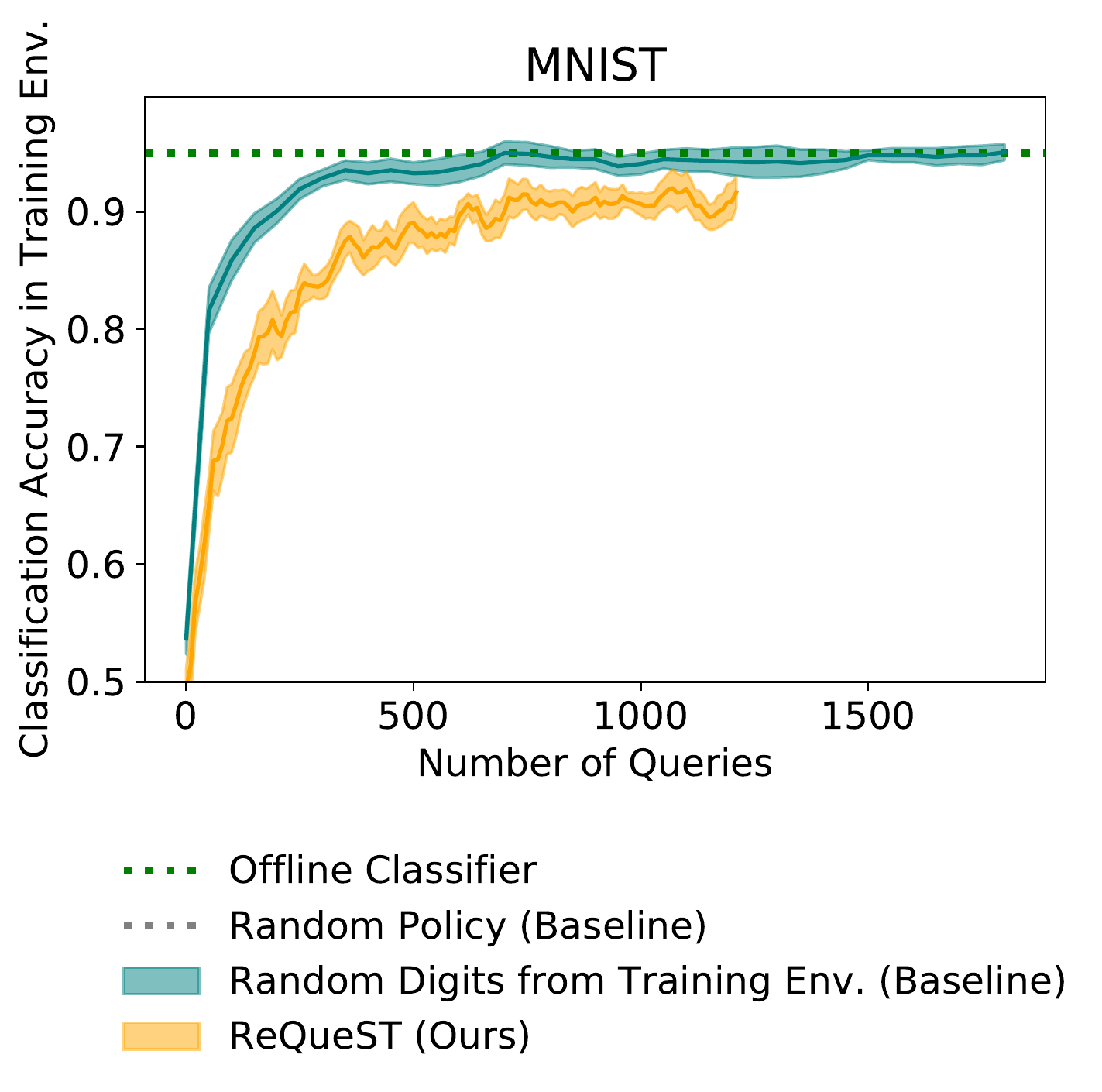}
    \caption{Our method performs worse than or comparably to the baselines in Section \ref{exp-robustness}, when the reward model is evaluated in the training environment instead of the test environment. Since there is no state distribution shift in this setting, training on real trajectories from the training environment (baselines) is more effective than training on hypothetical trajectories synthesized using our method (ReQueST). We do not include results for Car Racing, since the test environment is already identical to the training environment in this domain.}
    \label{fig:trainenvres}
\end{figure}

\begin{figure}
    \centering
    \includegraphics[width=\linewidth]{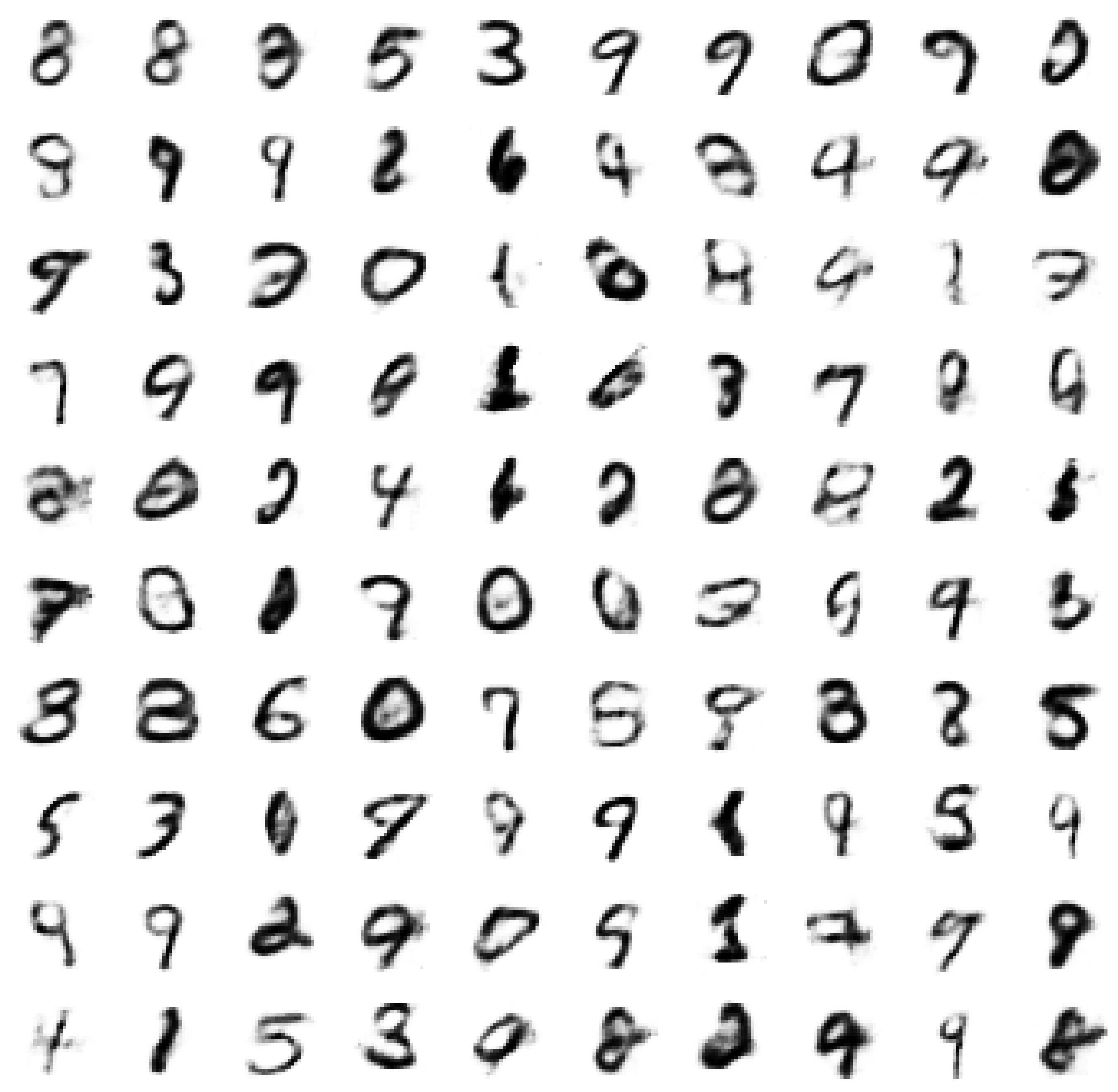}
    \includegraphics[width=\linewidth]{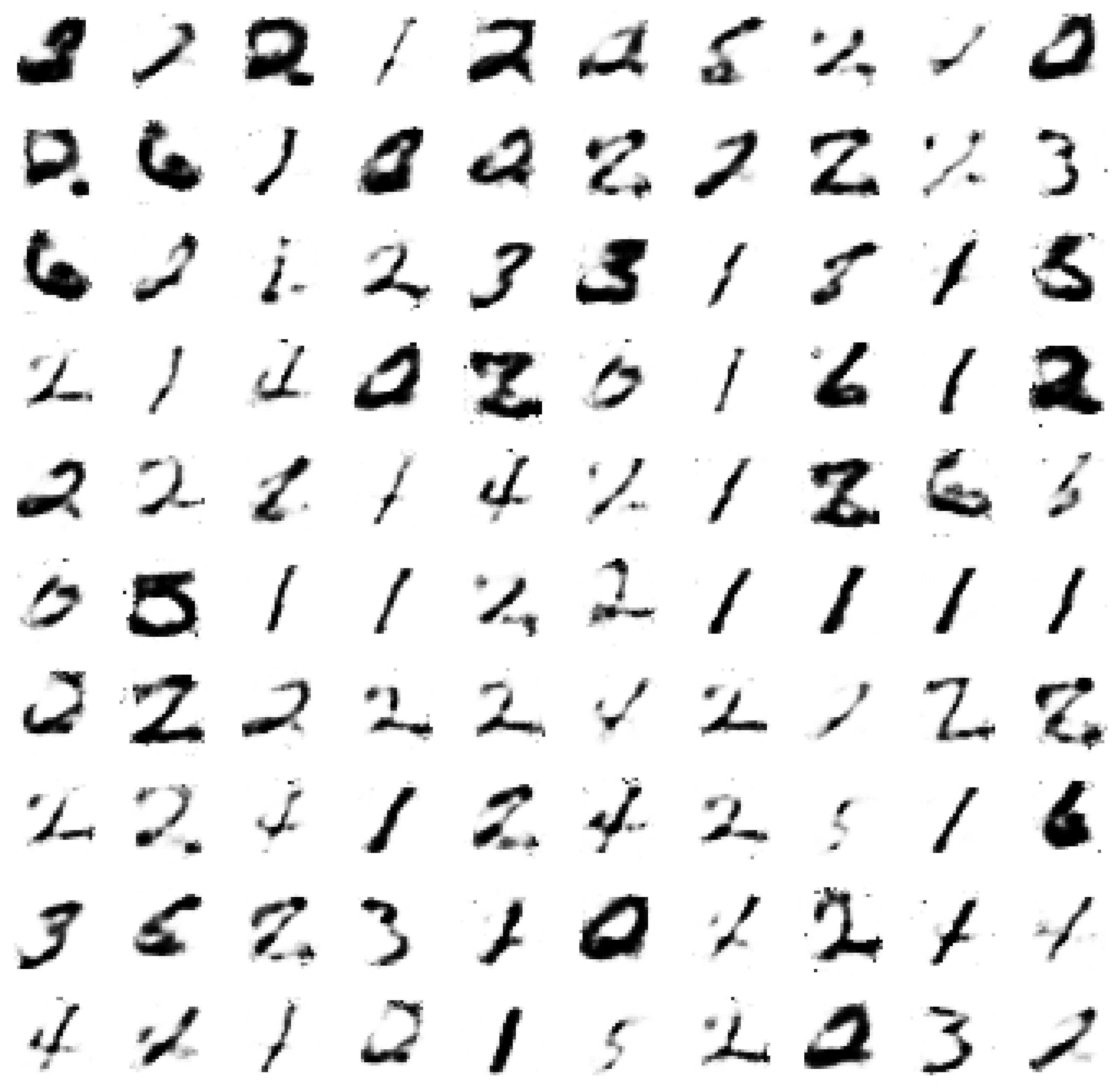}
    \caption{Examples of MNIST queries that optimize different AFs, illustrating the qualitative differences in the hypotheticals targeted by each AF. Top 10 rows: uncertainty-maximizing queries. Bottom 10 rows: novelty-maximizing queries. The uncertainty-maximizing queries are digits that appear ambiguous but coherent, while the novelty-maximizing queries tend to cluster around a small subset of the digits and appear grainy.}
    \label{fig:mnist}
\end{figure}

\begin{figure}
    \centering
    \includegraphics[width=0.85\linewidth]{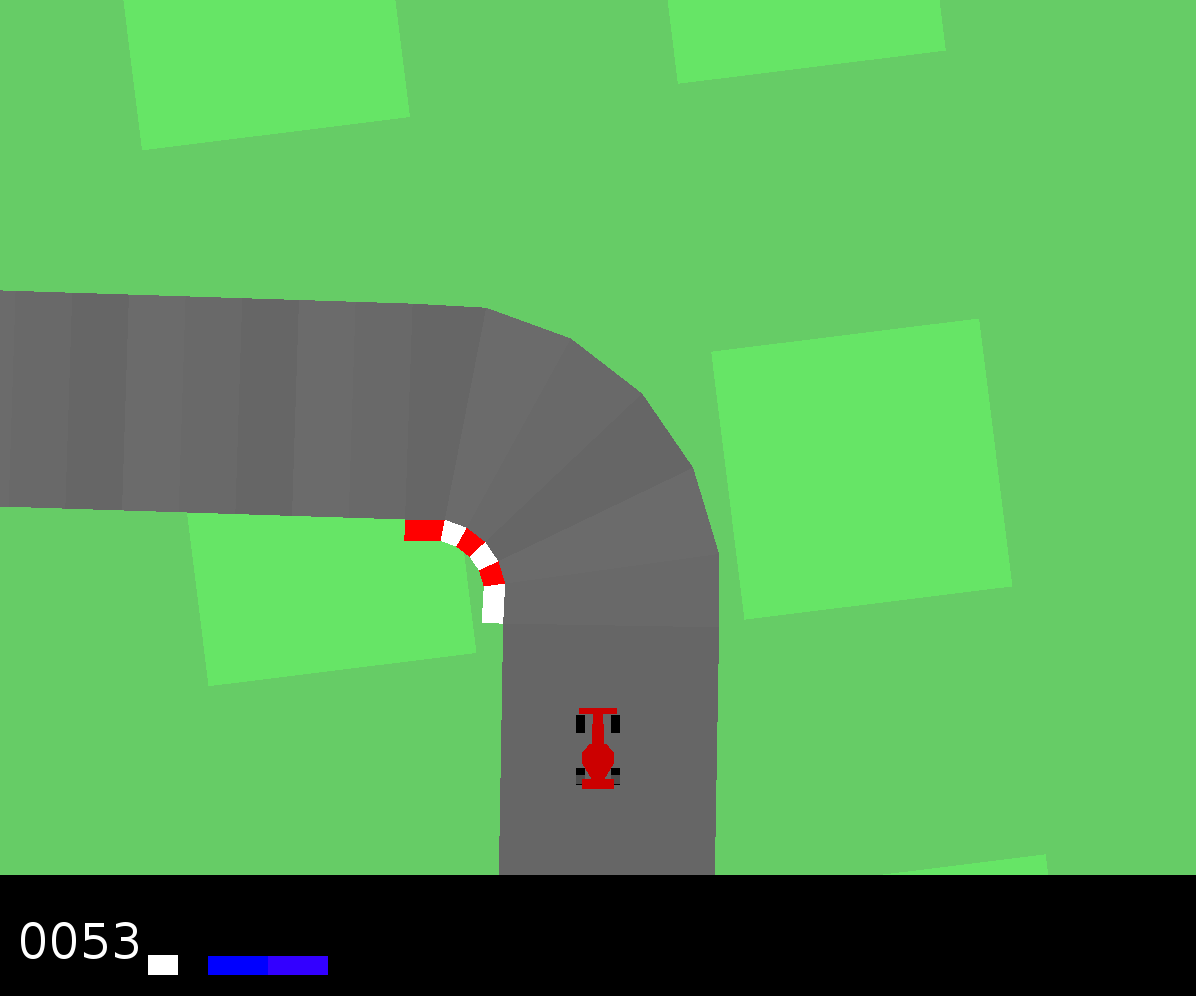}
    \caption{A screenshot of the image-based Car Racing video game in the OpenAI Gym.}
    \label{fig:carracing-screenshot}
\end{figure}

\begin{figure*}[t]
    \centering
    \includegraphics[width=\linewidth]{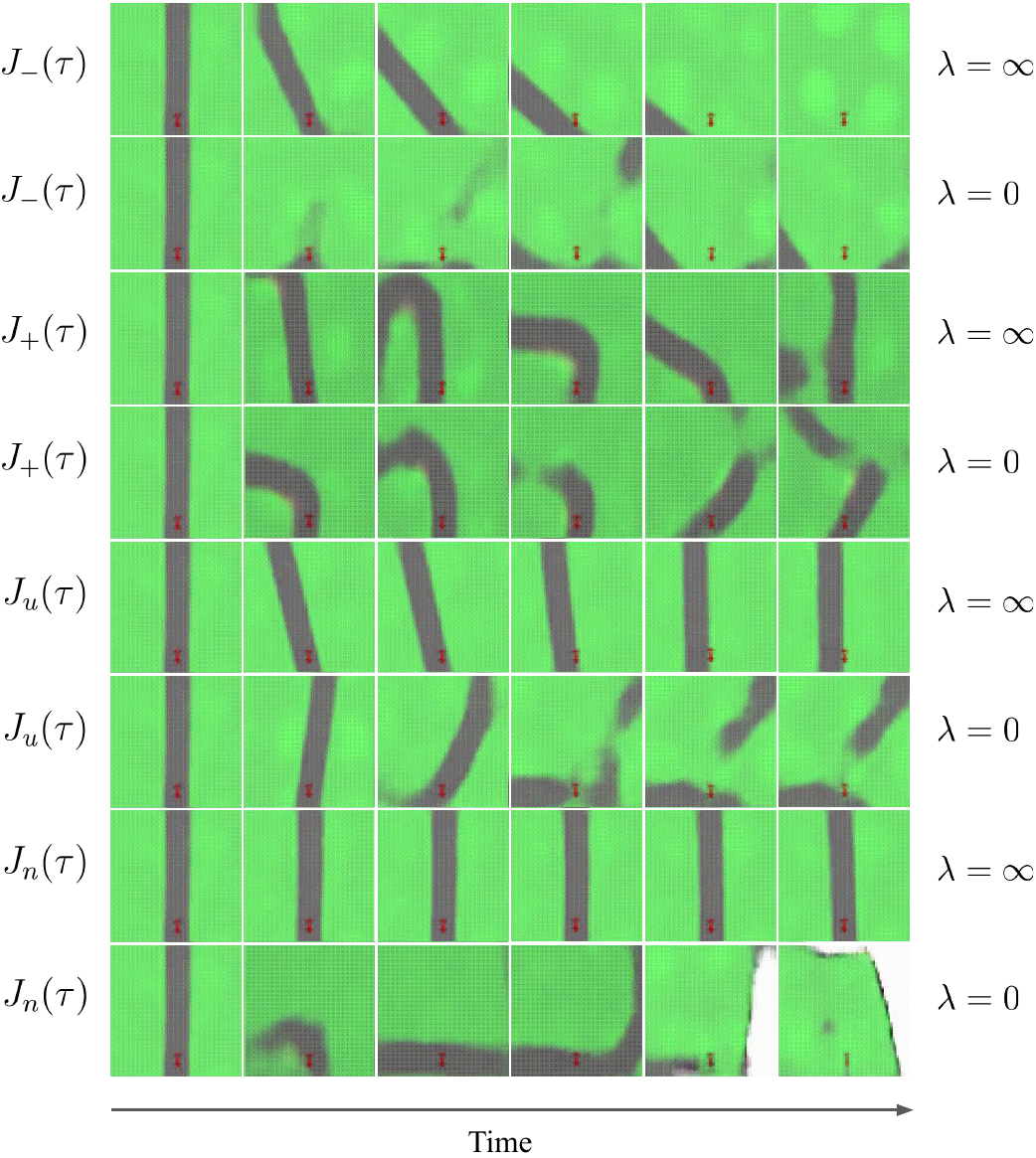}
    \caption{Examples of Car Racing queries that optimize different AFs with different settings of the regularization constant $\lambda$, illustrating the qualitative differences in the hypotheticals targeted by each AF, and the trade-off between producing realistic ($\lambda = \infty$) vs. informative ($\lambda = 0$) queries. When $\lambda = \infty$, the reward-maximizing query shows the car driving down the road and making a turn; the reward-minimizing query shows the car going off-road as quickly as possible; the uncertainty-maximizing query shows the car driving to the edge of the road and slowing down; and the novelty-maximizing query shows the car staying still, which makes sense since the training data tends to contain mostly trajectories of the car in motion. When $\lambda = 0$, most of the behaviors are qualitatively similar to their $\lambda = \infty$ counterparts, but less realistic and more aggressively optimizing the AF -- only the novelty-maximizing query is qualitatively different, in that it seeks the boundaries of the map (the white void) instead of staying still. Full videos available at \url{https://sites.google.com/berkeley.edu/request}.}
    \label{fig:carracing}
\end{figure*}

\end{document}